%% file: main.tex
\documentclass[acmsmall,screen,review=FALSE, anonymous=FALSE]{acmart}

\usepackage{graphicx}
\usepackage{tcolorbox}
\usepackage{tikz}
\usepackage{tikz-qtree}
\usetikzlibrary{trees}
\usepackage{multirow}
\usepackage{multicol}
\usepackage{array}

\AtBeginDocument{%
  \providecommand\BibTeX{{%
    \normalfont B\kern-0.5em{\scshape i\kern-0.25em b}\kern-0.8em\TeX}}}

\setcopyright{acmcopyright}
\acmJournal{PACMHCI}
\acmYear{2022}
\acmVolume{6}
\acmNumber{CSCW2}
\usepackage{csquotes}

\begin{document}

\title{Documenting Data Production Processes: A Participatory Approach for Data Work}

\author{Milagros Miceli}
\email{m.miceli@tu-berlin.de}
\orcid{0000-0003-0585-3072}
\affiliation{
 \institution{DAIR Institute, Technische Universit{\"a}t Berlin, and Weizenbaum Institute}
 \city{Berlin}
 \country{Germany}
}

\author{Tianling Yang}
\email{tianling.yang@tu-berlin.de}
\orcid{0000-0002-3746-7814}
\affiliation{%
\institution{Technische Universit{\"a}t Berlin \& Weizenbaum Institute}
  \city{Berlin}
  \country{Germany}
}

\author{Adriana Alvarado Garcia}
\email{adriana.ag@ibm.com}
\orcid{0000-0002-4230-3777}
\affiliation{
  \institution{IBM Thomas J. Watson Research Center}
  \city{Yorktown Heights, NY}
  \country{United States}
  }
 \additionalaffiliation{Georgia Institute of Technology, United States}

\author{Julian Posada}
\email{julian.posada@yale.edu}
\orcid{0000-0002-3285-6503}
\affiliation{%
 \institution{Yale University}
 \city{New Haven, CT}
 \country{United States}}
\additionalaffiliation{University of Toronto, Canada}

\author{Sonja Mei Wang}
\email{s.wang.1@tu-berlin.de}
\orcid{0000-0003-1219-8117}
\affiliation{
\institution{Technische Universit{\"a}t Berlin}
\city{Berlin}
\country{Germany}
}

\author{Marc Pohl}
\email{}
\orcid{}
\affiliation{
\institution{Independent Researcher}
 \city{Berlin}
  \country{Germany}
}

\author{Alex Hanna}
\email{}
\orcid{}
\affiliation{
\institution{DAIR Institute}
\country{United States}
}

\renewcommand{\shortauthors}{Miceli, et al.}

\begin{abstract}

The opacity of machine learning data is a significant threat to ethical data work and intelligible systems. Previous research has addressed this issue by proposing standardized checklists to document datasets. This paper expands that field of inquiry by proposing a shift of perspective: from documenting datasets toward documenting data production. We draw on participatory design and collaborate with data workers at two companies located in Bulgaria and Argentina, where the collection and annotation of data for machine learning are outsourced. Our investigation comprises 2.5 years of research, including 33 semi-structured interviews, five co-design workshops, the development of prototypes, and several feedback instances with participants. We identify key challenges and requirements related to the integration of documentation practices in real-world data production scenarios. Our findings comprise important design considerations and highlight the value of designing data documentation based on the needs of data workers. We argue that a view of documentation as a boundary object, i.e., an object that can be used differently across organizations and teams but holds enough immutable content to maintain integrity, can be useful when designing documentation to retrieve heterogeneous, often distributed, contexts of data production. 


\end{abstract}

\ccsdesc[500]{Human-centered computing~Empirical studies in collaborative and social computing}
\ccsdesc[300]{Human-centered computing~Collaborative and social computing design and evaluation methods}


\keywords{dataset documentation, data production, data work, data labeling, data annotation, machine learning, transparency}

\maketitle

\section{Introduction}

In recent years, research has turned its attention \textcolor{black}{toward} the datasets used to train and validate machine learning (ML) models \cite{scheuerman2021, denton2021, Sambasivan2021, paullada2020, crawford2019, sen2015, scheuerman2020}.  Work in this area \cite{bender2018, gebru2020, geiger2020, holland2018, afzal2021} has proposed frameworks to document and disclose datasets’ origins, purpose, and characteristics to increase transparency,  help understand models’ functioning, and anticipate ethical issues comprised in data.  Frameworks such as Datasheets for Datasets \cite{gebru2020}, Dataset Nutrition Label \cite{holland2018}, and Data Statements \cite{bender2018} have become increasingly influential. 
These forms of documentation are useful for providing a snapshot of a dataset’s state at a given moment. Still, one persistent challenge is that datasets are living artifacts \cite{balayn2021, jo2020, scheuerman2021} that evolve and change over time. Moreover, their production involves actors with varying amounts of decision-making power.  Capturing such variations requires, as we will argue, a shift of perspective: \emph{from documenting datasets to documenting data production processes.}

In this paper, we understand \emph{documentation} as a tool and the process of making explicit the contexts, actors, and practices comprised in ML data work, as well as the relationships among these elements. 
With the term \emph{data work}, we refer to tasks that involve the generation, annotation, and verification of data \cite{miceli2022}.  We understand \emph{data} as a process \cite{muller2022}, not a fact, and lean on CSCW and HCI work that has focused on data work in the ML pipeline \cite{ Irani2015,Sambasivan2022,Sambasivan2021, miceli2020}, studying the socio-economic situation of data workers \cite{hara2018, fan2020, ross2010} and the power relations involved in the labor process \cite{irani2013, salehi2015a, martin2014}.  Moreover, this paper expands our previous work \cite{miceli2020, miceli2021, miceli2022Dispositif} that studies a segment of data production that is outsourced through business process outsourcing companies (BPOs) that employ data workers from marginalized populations, mostly in developing countries.  Unlike crowdsourcing platforms, where algorithms manage the labor process \cite{Stark2021,Wood2018c,Rosenblat2016}, BPOs are characterized by traditional and localized managerial structures and business hierarchies \cite{miceli2022Dispositif}.  

Building on that work, the focus of the present paper is exploring ways of \emph{making distributed processes of data production more reflexive and participatory through documentation.} Promoting reflexivity through documentation means fostering collaborative documentation practices, including feedback loops, iterations, and the co-creation of tasks. Such considerations can help produce documentation that serves to interrogate taken-for-granted classifications and hierarchies in data work \cite{miceli2021}.

We report on 2.5 years of research comprising 33 semi-structured interviews, five co-design workshops, and several feedback instances where preliminary findings and prototypes were discussed with participants.  We draw on participatory design to work closely with two BPO companies located in Argentina and Bulgaria. We focus on the perspectives of data workers at both research sites and identify \textcolor{black}{the} challenges and requirements of integrating documentation in real-world data production scenarios. 

To that end, our research questions are:
\begin{enumerate}
    \item How can documentation reflect the iterative and collaborative nature of data production processes?
    \item How can documentation practices contribute to mitigating the information asymmetries present in data production and ML supply chains? 
    \item What information do data workers need to perform their tasks in the best conditions, and how can it be included in the documentation?
\end{enumerate}

Our findings show how the participants prioritized documentation that is collaborative and circular, i.e., documentation that is able to transport information about tasks, teams, and payment to data workers and communicate workers' feedback back to the requesters.  We discuss design considerations related to the integration of documentation practices as integral part of data production, questions of access and trust, and the differentiation between the creation and use of documentation. We reflect on these observations and the ideas of our participants through the acknowledgment that documentation is a \textit{boundary object} \cite{star1989,star2010}, i.e., an artifact that can be used differently across organizations and teams but holds enough immutable content to maintain integrity.   We further discuss a series of design and research implications that the acknowledgment of documentation as a form of boundary object entails.

This investigation extends previous work on dataset documentation in two aspects. 
First, instead of documenting datasets, we focus on the \emph{documentation of data production processes}, particularly the intra- and inter-organizational coordination comprised in data production, taking into account the iterative character of such collaboration and its impacts on data practices and datasets. This type of documentation can help break with patterns of information gatekeeping and top-down communication in data production by creating documentation that enables and reflects feedback loops and iterations.
Second, our investigation aims at producing \emph{documentation for and with data workers.} In this sense, we explore documentation forms that enable data workers' participation in the discussion of data production processes. Through this approach, we aim to open up paths for designing documentation that captures the needs and practices of data workers in contexts where their voices and labor tend to remain silenced \cite{irani2013, martin2014, miceli2020, miceli2022, salehi2015a,tubaro2019, gray2019, Posada2022}.


\section{Related Work}

\subsection{The Documentation of ML Datasets and Data Pipelines}

What counts as documentation is linked to the disciplinary background of each research community. For instance, while the field of open science envisions the documentation of data provenance and protocols \textcolor{black}{toward} reproducibility \cite{gil2016,goodman2014}, the Fairness, Accountability, and Transparency in \textcolor{black}{the} machine learning community is concerned with documenting ML models to prevent bias and improve transparency \cite{mitchell2019}. In this paper, we approach the documentation of data production by leaning on previous work that has advocated for transparency by proposing innovative frameworks for documenting ML datasets \cite{bender2018, gebru2020, geiger2020, holland2018, afzal2021}. These frameworks vary in the forms of documentation, such as datasheets, dataset nutrition labels, and data statements, and have different prioritized goals and intended field of use in their design. They share similarities in requiring a standardized set of information for documenting the motivation \cite{gebru2020}, curation rationale \cite{bender2018}, composition \cite{gebru2020}, provenance \cite{bender2018, gebru2020, holland2018, afzal2021}, collection process and context \cite{bender2018, gebru2020}, actors involved \cite{bender2018, geiger2020} and tools used \cite{gebru2020, geiger2020}. Some of these groundbreaking frameworks have been successfully integrated and implemented in corporations such as Google and IBM as pilot projects \cite{zotero-12094, zotero-12092}, and several ML conferences that now require new dataset submissions \textbf{to be} accompanied by a datasheet, as evidenced by the recent NeurIPS Dataset and Benchmark track \cite{zotero-12093}.

The foregoing frameworks focus on the documentation of \emph{the datasets themselves}, offering a useful snapshot of the dataset's current state to dataset producers, users, and researchers. However, as previous work \cite{miceli2022} has argued, two fundamental aspects of ML datasets represent a challenge to such standardized documentation frameworks: (1) The collaborative nature of data work and data production \cite{passi2018, miceli2020,muller2021}, including the often messy relationship between stakeholders \cite{miceli2021}, which challenges the strict differentiation between data subjects, data users, and data producers assumed by standardized documentation frameworks; 
and (2) the fact that datasets are not static entities but rather evolving projects with flexible boundaries \cite{balayn2021, scheuerman2021}. 

In view of these challenges, recent research in data documentation has raised attention \textcolor{black}{toward} a consideration of data production pipelines, including the multiplicity of workflows and actors involved in producing ML datasets \cite{miceli2021, Sambasivan2022, denton2020, balayn2021, pushkarna2022}.  \citet{hutchinson2021a} put forward a set of questions structured by interconnected stages of the data life cycle. Conceiving this cycle as non-linear, the authors frame datasets as technical infrastructures and their production processes as “goal-oriented engineering,” which favors adopting methodologies with deliberation and intentionality to improve transparency and accountability over requiring only post-hoc justification. 
Relatedly, \citet{balayn2021} introduce a process-oriented lens to dataset documentation and expand the focus to the design of data pipelines and the reasoning behind them. This process lens can help identify and locate knowledge gaps in the pipeline and uncover places and procedures that are worth more rigorous documentation and careful reflection on their potential impact and harm.
Other studies  \cite{jo2020, famularo2021, scheuerman2021} have called for the professionalization of data work and care (i.e., the careful consideration of the dataset in terms of “the domain setting where it originates, and the potential questions modeling that data might answer or problems it might solve” \cite{wolf2019b}), the publication of data documentation, the implementation of institutional frameworks and procedures to promote documentation, and the consideration of data work as an own subfield of research \cite{jo2020, Sambasivan2021, scheuerman2021}. 

These investigations shift the perspective to regard datasets as living artifacts that are formalized by actors with different positions and decision-making power.  From this perspective, documentation could capture data pipelines and production contexts as well as various actors that shape datasets. At its ideal state, documentation should also be able to foster “a documented culture of reflexivity about values in their [data] work” \cite{scheuerman2021}. As goals, interests, and needs vary across different contexts and organizations, documentation should be flexible to accommodate local specificities and constraints and, at the same time, keep certain common identities. In line with \citet{pushkarna2022}, whose work in data documentation reflects the need for disclosure documents to be responsive to a larger body of stakeholders, including annotators, data curators, and users, we argue that recognizing documentation as a \emph{boundary object} can be helpful to approach and reflect on such challenge.

\subsection{Boundary Objects}

The notion of \textit{boundary object}, defined as “objects that both inhabit several communities of practice and satisfy the informational requirements of each of them” \cite{bowker1999}, is a useful lens to study how documentation operates across diverse contexts in the ML supply chain.
Not only do boundary objects facilitate communication among different groups, \textcolor{black}{but} they also translate and transfer knowledge to be accessible to all potential stakeholders. This way, boundary objects can meet stakeholders' specific information needs and work requirements, but they do not require them to have a panoramic grasp and full comprehension of every detail. The variety of maps of a city is an example of \textcolor{black}{a} boundary object. The maps have the same geographical boundary, namely the city, but different groups use them differently. For example, consumption-minded citizens may search information about local parks and restaurants, physical geographers may pay attention to mountains and rivers, botanists to certain horticultural areas and botanical gardens, and administrators may keep an eye on political boundaries. \citet{suchman2002} emphasizes the appropriability of objects into use contexts and working practices.
In her analysis of a civil engineering project, the author shows how heterogeneous artifacts were brought together and integrated into everyday work practices \cite{suchman2000}.
When such “artful integration” \cite{suchman2002} involves continuous and relatively stable relationships among different communities and shared objects jointly produced by them, "then boundary objects arise"\cite{bowker1999}.

 Within CSCW and HCI, there is a longstanding use of the boundary object notion to investigate how artifacts can be used across varying contexts and to facilitate cooperation and collaboration, such as in the medical order \cite{zhou2011}, in the coordination among different actors in  hospitals \cite{bossen2012, bossen2014}, in aircraft technical support \cite{lutters2002, lutters2007}, in disease management at home \cite{aarhus2010}, and in issue tracking systems \cite{bertram2010}.
This notion is also used to study organizational memory \cite{ackerman1998} and common information space \cite{bannon1997, schmidt2008}. 
\citet{lee2005, lee2007} further coins the term \emph{boundary-negotiating artifacts} to explore non-standardized, fluid informal objects that not only travel across but also destabilize and push boundaries. Moreover, some participatory design approaches consider prototypes and models as boundary objects that facilitate the exchange of information and knowledge, and make it possible to involve different stakeholders in the design process by establishing a shared syntax or
language \cite{pelle_2008, routledge_tools, carlile_2002, Haskel2017ParticipatoryDA}.

Socio-material perspectives point out that the features of physical objects impact the ways they behave and operate as boundary objects \cite{mcleod2010, huvila2011, hawkins2017}. Documents, as a specific form of boundary objects, can become part of communicative practices in three ways: they can be an object of evaluation, a medium to facilitate communication, and/or the arena where communicative practices are shaped \cite{osterlund2008, mcleod2010, huvila2011}.  Contexts of data production, where actors often pursue different goals and interests, pose their own challenges in terms of fostering shared values and effective communication and, finally, promoting common documentation practices. Given these challenges, documentation should adapt to the needs of different stakeholders while remaining stable enough not to lose its function across different settings. The acknowledgment of data documentation as a boundary object \cite{pushkarna2022} can provide a useful lens to approach and address some of the tensions arising from coordinating and collaborating across actors and organizations in data production. Specifically, the boundary object notion can add to existing work on data documentation in two ways. First, the acknowledgement of data documentation as a boundary object helps to make visible and specify the variety of actors involved in data production contexts with their own priorities, workflows, and needs. Secondly, this acknowledgment can add a practice-oriented perspective to the formal standardization of documentation frameworks and make work more visible. This includes work not only in the accommodations of standardized documentation frameworks to local needs and specificities, but also in the adaptation and improvement of existing work practices and workflows. Looking at documentation through the lens of boundary objects can be helpful \textcolor{black}{in fostering} documentation practices and design documentation frameworks that capture the practical, economic, and organizational decisions involved in data production and encoded in datasets.

\section{Method}

\begin{figure}[ht]
    \centering
    \includegraphics[scale=0.66]{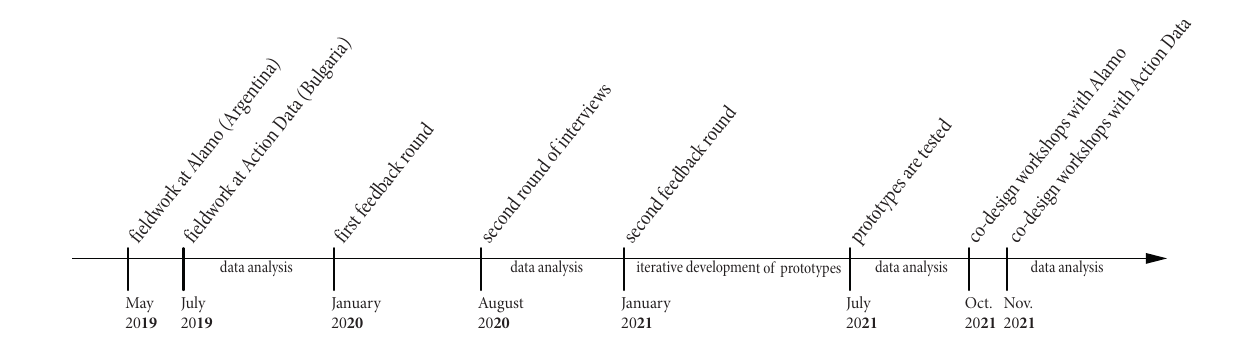}
    \caption{Timeline of the iterative phases of data collection and analysis comprised in this research.}
    \label{fig:timeline}
\end{figure}

This investigation comprises 2.5 years of research in collaboration with two BPO companies located in Buenos Aires, Argentina (Alamo), and Sofia, Bulgaria (Action Data), \textcolor{black}{specializing} in the collection and annotation of data for ML.  Our engagement with both organizations included fieldwork at their offices, interviewing phases, moments of feedback and discussion, the collaborative development of prototypes, and a series of co-design workshops. In this section, we provide details of our engagements with the participants and the methods used. Figure \ref{fig:timeline} shows how the research process evolved over time.

\subsection{Participatory Design}
\label{sec:PD}

At the core of Participatory Design (PD) is the active involvement of the communities that will use technologies and systems in their design. Moreover, PD recognizes the importance of the context in which technologies are used and the situated practices within that context \cite{routledge_introduction}.
In CSCW and HCI, participatory design methodologies have been deployed to explore various fields of inquiry, ranging from public services \cite{Corbett_2019, Sandjar_2019, Asad_2017, saxena_public_service_2020, flugge_pes_2021, costantino_library_2014, scott2022}, to healthcare \cite{kusunoki_2015, pina_2017, rothmann_2016}, to organizational settings \cite{madaio_2020, Woodruff_2018, Wolf_2020}. Additionally, PD has been used in the field of ML Fairness, Accountability, and Transparency to guide algorithmic accountability interventions \cite{Kattel2020}, learn about the concerns of communities affected by algorithmic decision-making systems \cite{Brown2019, Robertson2020}, and elucidate causal pathways to be input into complex modeling \cite{martin2020participatory}. Regardless of the discipline, what these previous efforts have in common is their commitment to building reciprocal partnerships to inform the design of systems and technologies based on the values of justice and participation. 

In this investigation, we use PD to explore the development, implementation, and use of documentation in real-world data production scenarios that involve many stakeholders that are geographically distributed. Similar to \citet{villacres_2020}, we understand participatory design as a long-term iterative process in which exploring participants' insights is as essential as tangible design outcomes. We use PD and co-design as synonyms \cite{costanza-chock2020}. 

Throughout the research and design process, we focused on three core elements of PD: \emph{having a say}, \emph{mutual learning}, and \emph{co-realization} \cite{routledge_methods, muller_thirdspace_2007}. We were intentional about listening to what the different actors involved in data production (the data workers, their managers, and the clients) had to say about dataset documentation in order to derive design considerations that would adapt to their needs. As the investigation evolved, we especially focused on the experiences of data workers because we realized they were the ones that usually \emph{did not have a say} in how workflows and processes are designed in ML supply chains. We discuss this shift of perspective in Section \ref{sec:findings}. 

While working on the research design, we put effort in remaining sensitive to the “politics of participation” \cite{costanza-chock2020} by trying to anticipate power asymmetries that could emerge among participants and between researchers and participants.  In this sense, the interviewers and workshop facilitators were at all times intentional about \textit{making room} for data workers' voices, for instance, by insisting on conducting in-depth interviews with them or \textcolor{black}{preventing} managers \textcolor{black}{from hogging} the mic during the workshops. Moreover, \citet{bratteteig_cscw_2016} argue that “language is power” and by speaking one's own language, discourse can be expanded and \textit{model monopoly}, i.e., the adoption of external mental models, can be avoided \cite{routledge_methods}. In this sense, it was important to us to let the participants maintain their preferred language\,---\,both verbally and visually. To that end, we provided interpretation and facilitation in the participants languages and worked largely with visual symbols \cite{bratteteig_cscw_2016}. In section \ref{sec:resimp}, we reflect upon our research and highlight a few implications for participatory research in terms of power, negotiation, and participation in corporate settings \cite{sloane2020}.

\subsection{The Participating Organizations}

Two companies dedicated to the collection and labeling of ML data participated in this investigation:

\begin{itemize}
    \item \emph{Alamo} (pseudonym), located in Buenos Aires, Argentina.  
    \item \emph{Action Data} (pseudonym), located in Sofia, Bulgaria.
\end{itemize}

Both BPOs are impact sourcing companies. \textit{Impact sourcing} refers to a branch of the outsourcing industry that proposes employing workers from poor and marginalized populations with the twofold aim of offering them a chance in the labor market and providing information-based services at lower prices. In both organizations, data workers perform tasks related to the collection and annotation of data. These tasks are requested by external clients located in various regions of the world, who seek to outsource the production of datasets used to train and validate ML models. The relationship between requesters and data workers is mediated by managers, reviewers, and quality assurance (QA) analysts that form a hierarchical structure where data workers occupy the lowest layer \cite{miceli2020}.

\subsection{Data Collection}

\subsubsection{Semi-Structured Interviews}
\hspace{1cm}

A total of 33 interviews were included in the present investigation. 19 were conducted in person in 2019. Due to the on-going COVID-19 pandemic and the resulting travel restrictions, 14 interviews were conducted remotely between August 2020 and September 2021. 

\input{participants}

All interviews were semi-structured. Those conducted in 2019 were in-depth interviews that explored the lived experiences of data workers and aimed at understanding the contexts in which data work is carried out \cite{miceli2020}. The average duration of those interviews was 65 minutes. For the analysis comprised in this paper, we selected only the in-depth interviews that revealed insights about documentation practices. In 2020 and 2021, we carried out a second round of interviews focused on the topic of documentation. Each one of those interviews had an average duration of 50 minutes. We interviewed actors who were actively involved in the documentation of projects at both companies. In addition, we interviewed machine learning practitioners with experience outsourcing data work to platforms and BPOs. Some of these practitioners were direct clients of Alamo and Action Data. Table \ref{tab:participants} provides an overview of the interviews and participants included in this work.

The interviews allowed us to explore documentation practices carried out in different organizations and identify perceived challenges, experiences, and needs. Some interview partners were interviewed more than once at different stages of the design process. For each interview, the informants received a gift card for €30 (or the equivalent in their local currency) for an online marketplace. The audio of the interviews was recorded and, later on, transcribed. The name of the interview partners and any other elements that would help identify them were changed. The names that appear in the excerpts included in this paper are pseudonyms that the interview partners chose for themselves.

\subsubsection{Co-Design Workshops}
\label{sec:METHworkshops}
\hspace{1cm}

Given the multiplicity of actors involved in processes of data production \cite{miceli2020,miceli2021}, we saw value in inviting data workers, their managers, and their clients to a series of co-design workshops to discuss documentation practices. Table \ref{tab:participants} offers an overview of the organizations and individuals participating in each workshop series. 

The workshop series with each company took place on different days with a similar structure but was tailored to each setting and based on the ongoing projects and clients of the respective BPO. Due to the COVID-19 pandemic, all sessions took place online via Zoom and the activities were conducted in parallel on the visual tool Miro. As compensation, all workshop participants received €15 per hour of participation. The sessions were video-recorded and, later, transcribed. In addition, note-takers registered the activities and interactions and one of the authors produced a real-time graphic recording on Miro. The graphic recording was kept visible to participants as it evolved throughout the workshop sessions. 

\begin{figure}[ht]
    \centering
    \includegraphics[scale=0.30]{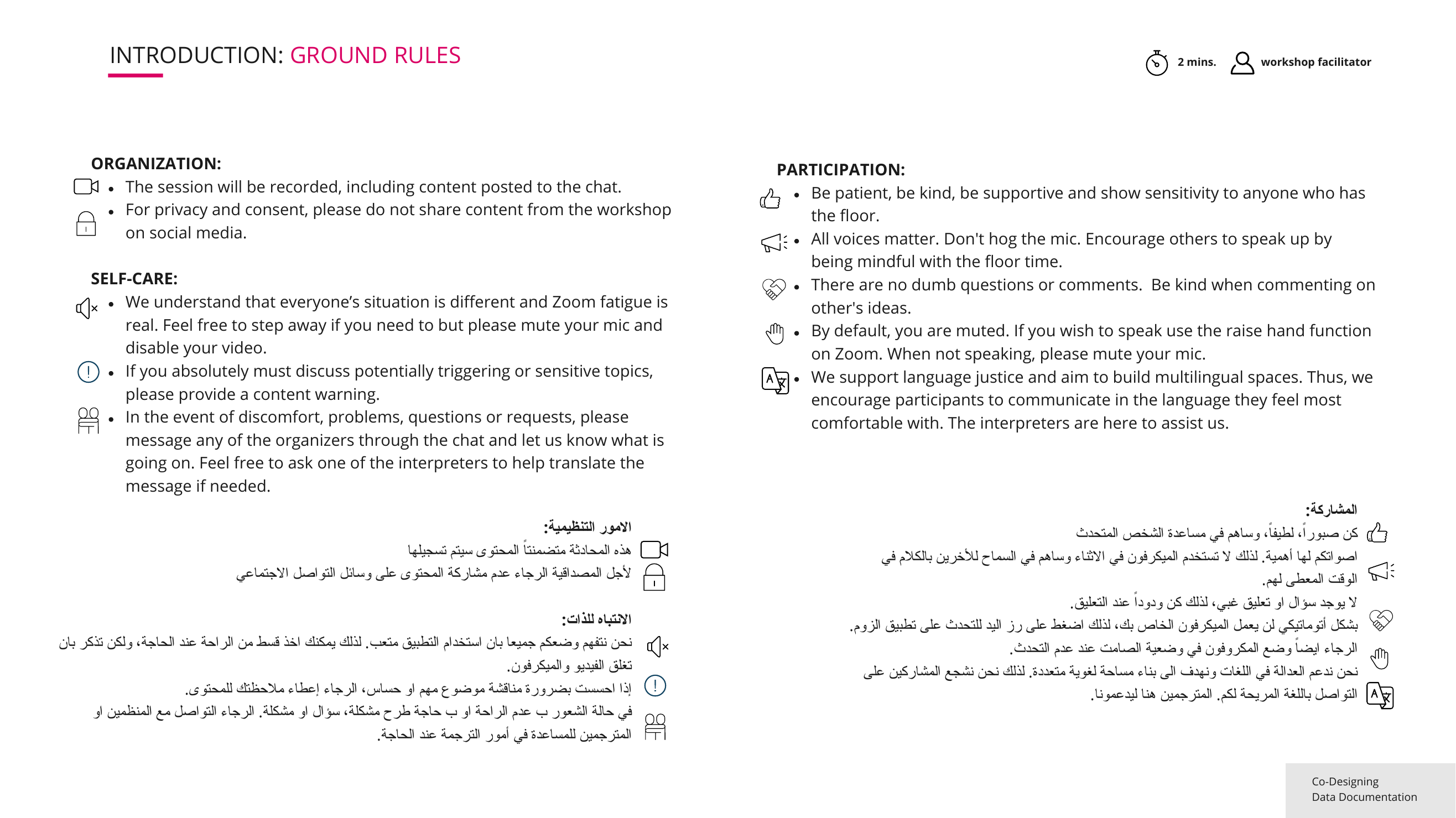}
    \caption{The ground rules for participation were visible at all times throughout the co-design sessions with Action Data's workers. A Spanish version of the same rules was introduced at the workshops with Alamo.}
    \label{fig:groundrules}
\end{figure}

The workshops with the Argentine company Alamo were conducted in Spanish, the native language of both participants and the co-authors who served as facilitators.
The sessions took place on October 26 and 28, 2021. Each session comprised two hours. 17 participants attended the workshops, including four managers and 13 data workers. We suggested Alamo's management include one of the company's clients in the workshops but this request was repeatedly declined. Our proposal to include a third workshop session for data workers only (without the presence of management) was also declined due to scheduling issues according to the company’s management.

The 3-day workshop series that we organized with Action Data took place on November 3, 5, and 8, 2021. On day 1, we hosted a 90-minute session and invited Action Data's management (two participants) located in Bulgaria and 13 of the data workers that Action Data manages through its partner organization, Ranua, located in Syria.
For day 2, we prepared a 3-hour workshop and had three members of one of Action Data's client organizations \,---\,a Nordic university\,---\,join \textcolor{black}{in}. 
Finally, day 3 was conceived as a 2-hour round table only for the Syrian data workers to discuss their impressions without the presence of managers and clients. This was motivated by the acknowledgment that the presence of managers can influence the level of comfort workers feel when expressing their opinions. 
In all sessions, the co-authors who served as main workshop facilitators spoke in English with simultaneous interpretation. For the breakout rooms, we offered facilitation in English and Arabic, depending on the needs of each group. 

The power differentials present among the participant groups motivated many of our decisions in terms of how to design the respective sessions with Alamo and Action Data. The participation rules introduced at the beginning of the workshops reflect our stance (see Fig. \ref{fig:groundrules}). The acknowledgment of such power differentials also demanded much attention and flexibility on our side to quickly adapt and reformulate activities in view of difficulties identified while conducting the workshops. 

\subsection{Data Analysis}

The collected data were analyzed through reflexive thematic analysis (RTA) as developed by \citet{braun_clarke_thematic} and \citet{braun_2019_thematic}. 
RTA is a type of thematic analysis that places importance on researcher subjectivity and sees the researcher as playing an active role in the knowledge production process, a value which is emphasized by the word “reflexive.” We describe the position from which we have conducted this research in the next subsection, \ref{sec:positionality}.

\input{codebook.tex}

To analyze the data collected in the present investigation, two of the co-authors explored the interview and audio transcriptions first. They started with a first coding round after which two further authors revised and added codes. A meeting followed where the four coders exchanged impressions and discussed disagreements around the codes. Additionally, they revisited video recordings of the workshops as well as the  activities captured on Miro boards. 

We constructed candidate themes after revisiting coding categories. \citet{braun_2019_thematic} argue that themes "do not emerge fully-formed" and that there are first candidate themes that are developed from earlier phases, before settling on the final themes. It took us several iterations until arriving at a (more-or-less) final set of themes. Table \ref{tab:codes} provides an overview.

The transcriptions of the data collected at Alamo were in Spanish. The data collected at Action Data was in English, with the exception of three interviews and all workshop sessions that included English and Arabic-speaking participants. In those cases, the English interpretations as performed by the interpreters were transcribed. Coding the transcriptions in both Spanish and English was possible because all authors are proficient in English and three of them are native Spanish speakers. Some of the excerpts included in the Findings sections were originally in Spanish and were translated into English only after the analysis was completed. 

It is worth mentioning that some of the initial interview data had been previously coded for an earlier study conducted by some of the authors \cite{miceli2020} using constructivist grounded theory \cite{charmaz2006}, a method that was consistent with the exploratory spirit of the fieldwork and the first interviews conducted in 2019. For the present paper, we revisited some of those interviews and re-coded them along with the newly collected data using RTA. 


\subsection{Positionality Statement}
\label{sec:positionality}
Our ethical stance throughout this investigation was centered on respecting workers' expertise \cite{routledge_ethics, sanders_cocreation, costanza-chock2020} and creating spaces where they could describe work practices and conditions, challenges, and needs. A central motivation\,---\,and certainly a challenge\,---\,of this research was to empower data workers to speak up and participate in shaping their own workflows and tools \cite{steen_2007}. 

Our position and privilege vis-à-vis our study participants \textcolor{black}{have} been in constant interrogation and discussion throughout data collection, analysis, and while considering the implications of this investigation. Opting for an analysis method that would help us to acknowledge and make our positionalities as researchers explicit was essential to us. Following the idea of conducting a \emph{reflexive} thematic analysis, it seems appropriate to disclose some elements of the authors' backgrounds that might have informed the findings we will present in the next section. 

All authors except one are academics working in institutions located in the Global North. Most of us carry passports that are foreign to the places where we live and work. Four of us are sociologists, one a designer, and two are computer scientists. We all work in the field of Human-Computer Interaction and Critical Data Studies. Three of us are Latin American, two were born in China, one of us is German, and another is a first-generation US-American. All of us are first-generation academics. Our position as researchers living and working in the Global North provides us with privilege that our study participants do not hold. Despite being born and raised within working-class families, some of us in the same country as some of the participants and having too experienced migration, we acknowledge that our experiences differ from that of the participants in the sense, for instance, that we have never experienced war or had to flee our home countries to become refugees. This is \textcolor{black}{e}specially relevant vis-à-vis the Syrian data workers that participated in this investigation. The acknowledgment of such privilege has guided several of the decisions that we have made, for instance, in terms of mitigating power differentials in data collection. In this sense, we must mention that, despite our best efforts, some power asymmetries could not be mitigated and we could only acknowledge and reflect upon them during data analysis and in this paper (see Sections \ref{sec:METHworkshops} and \ref{sec:resimp}).

\section{Findings}
\label{sec:findings}


Our findings are presented in a way that reflects the diverse priorities, opinions, positions, and power of our participants. 
For each research site (Alamo in Sect. \ref{sec:Alamo} and Action Data in Sect. \ref{sec:ActionData}), we first provide details of our engagement with the respective partner organization and the preliminary knowledge obtained through the interviewing phases. We then move on to describe salient discussions that took place at the workshops. In Section \ref{sec:FindSum}, we present a summary of findings including commonalities and differences across sites. 

Long-term engagements with participants are key to participatory research \cite{vargas-cetina2020a, chibnik2020a} and, as \citet{sloane2020} argue, it is important “to make the tensions that characterize the goal of long-term participation in ML visible, acknowledging that partnerships and justice do not scale in frictionless ways, but require constant maintenance and articulation with existing social formations in new contexts.” We believe that there is a contribution to be made by placing our findings in conversation with the vicissitudes of conducting participatory research, especially with the inter-mediation of partner organizations. This is why we show how our observations evolved over 2.5 years and across sites and methods.  We are convinced that our findings would not be the same had we stopped after the interviews. In the same way, without a description of the interviewing process and how our relationship \textcolor{black}{with} the participant organizations evolved, the ideas emerging from the co-design workshops would lose some of their nuances.

\subsection{Case 1: Alamo}
\label{sec:Alamo}

The Buenos Aires-located company that we will call \textit{Alamo} is, at the time of this investigation, a medium-sized organization with around 400 data workers distributed in three offices located in Argentina and two other Latin American countries. Its main client is a large regional e-commerce corporation. 90\% of Alamo's projects are requested by this client and include image segmentation and labeling, data collection, and content moderation. The Buenos Aires office employs around 200 data workers, mainly young people living in very poor neighborhoods or slums in and around the city. Looking for workers in poor areas of Buenos Aires is part of Alamo's mission as an impact sourcing company. Alamo's data workers are offered a steady part- or full-time salary and benefits, which contrasts with the widespread contractor-based model observed in other BPOs and data work platforms \cite{Posada2020a}. Even so, Alamo's workers receive the minimum legal wage in Argentina (the equivalent to US\$1.50/hour in 2019) and their salaries situate them below the poverty line.

\subsubsection{Investigating Alamo's Documentation Practices}
\label{sec:AlamoInt}
\hspace{1cm}

\textcolor{black}{During} fieldwork in 2019, one of the observations that stood out was the company's stated commitment to fostering its workers' professional growth \cite{miceli2020}. In contrast with this mission, Alamo's emerging documentation practices seemed, at the beginning of this investigation, almost completely oriented \textcolor{black}{toward} surveilling workers and measuring their performance. Noah, one of the data workers that we interviewed in 2019 described that documentation process as follows: 

\begin{displayquote}
For instance, for this project, we have three metrics for individual performance and now we're about to add one more. Mostly, we quantify and document errors or rollbacks reported by the requester.
\end{displayquote}

Nancy, who was one of Alamo's project managers in 2019, explained what was done with the documented performance metrics:

\begin{displayquote}
Within the company's structure, QA [the quality assurance department] then takes those metrics and puts together something like a “scoreboard” that shows everyone's performance and that is visible \textcolor{black}{to} the whole company.
\end{displayquote}

After our first fieldwork period and once we had analyzed the collected data, we provided Alamo with a report summarizing our observations.  We found that there was a contradiction between the company's mission of fostering worker growth and its emphasis on quantifying individual performances, and suggested reorienting the purpose of documentation to include information that could help workers understand the background and purpose of their work. We have discussed those observations and their implications at length in previous work \cite{miceli2020, miceli2022Dispositif}. After presenting our observations, we engaged in a conversation with Alamo's co-founder Priscilla who, as shown by the following interview excerpt, agreed with our suggestions and expressed her desire \textcolor{black}{to establish} documentation practices based on the needs of data workers:

\begin{displayquote}
In my opinion, this is what we need to do.  Listen more to their needs and ask them what they need to perform better and do their work better. And we must generate and provide the information they need.  We need to make time for that.
\end{displayquote}

In September 2020, we resumed contact with Alamo's management and invited the company to participate in our investigation of data production documentation, starting with a series of follow-up interviews. By then, the company was already in the process of restructuring its documentation practices. Nati, one of Alamo's former QA analysts, had just been promoted to “continuous improvement analyst” and put in charge of bringing together the company's quality standards and its goals in terms of promoting workers' professional growth. Nati's main task was re-designing Alamo's documentation practices. She described this process as follows:

\begin{displayquote}
During the past year, we made an effort to ask everyone “what would you like to see reflected in documentation? What do you need?” After that, we started dropping old practices that didn't make sense anymore. We threw them away, and started anew. We learned a lot about empathy too, you know? This work is very much about empathy [...] because we assign workers in each team according to their competencies and their skills, and they need to know what projects are about to actually make the most out of those skills. 
\end{displayquote}

What came out of the process led by Nati was a series of documentation practices strongly oriented \textcolor{black}{toward} \emph{knowledge preservation}, i.e., the documentation of workflows, tools, specific details about the conduction of projects, best practices, and lessons learned \cite{miceli2021}. In Nati's words, documentation was envisioned as an educational resource for Alamo's workers and its aim was \textcolor{black}{to mitigate} information loss due to fast worker turn around. A special team led by Nati was put in charge of developing and implementing this vision. With the documentation guidelines and templates designed by them, each team leader would document the projects carried out by their data workers. Moreover, following one of our suggestions, the company saw value in documenting contextual information about the ML products that are trained or validated using data produced at Alamo. Management then started to seek clients' involvement in documentation practices. The first step was sharing the project documentation with the project's requester and asking for feedback. These collaborative practices progressed to the point that some clients became co-producers and users of the documentation produced by Alamo, as Pablo, leader of one of Alamo's data labeling teams described:

\begin{displayquote}
I mean, it depends on the client. Luckily this requester is very committed and is always looking at our documentation, providing feedback, checking if the information is up to date. Sometimes they just jump in and add information, they collaborate on the best practices manual. It's also good for them to see if their information is updated too, not just for us but for the requester's own team, because they share these documents within the requester's company.
\end{displayquote}

While this arrangement was beneficial for the clients, Alamo was providing the infrastructure and most of the workforce that made documentation possible. Alamo's workers started to feel the burden, as team lead Pablo described: 

\begin{displayquote}
We definitely noticed the change: from not documenting at all to documenting everything. It also depends on the project. For instance, our main client [a large regional e-commerce corporation] brought many more requirements for us to implement in our documentation in terms of what they wanted to see reflected. There are new documents and new guidelines all the time. 
\end{displayquote}

\begin{figure}[ht]
    \centering
    \includegraphics[scale=0.40]{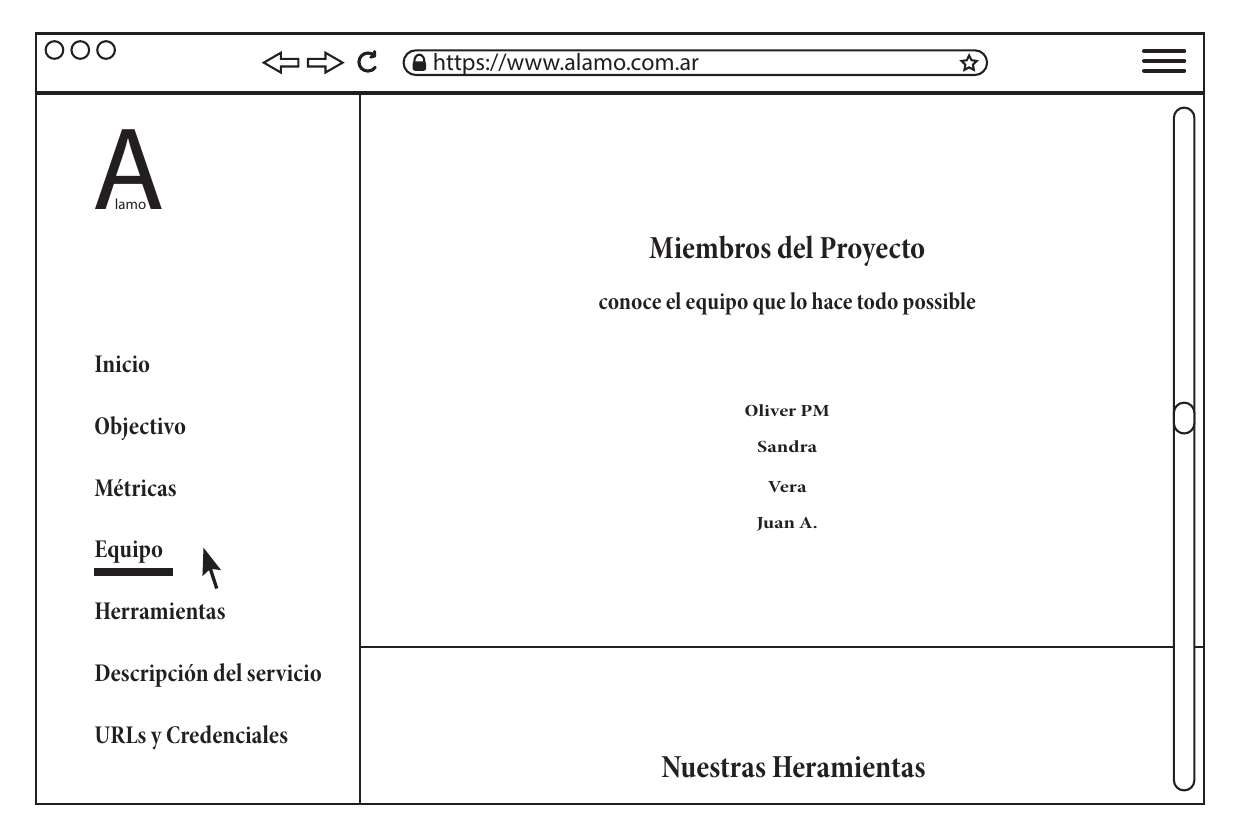}
    \caption{The wiki-based documentation prototype created by Alamo to brief and train workers.}
    \label{fig:wiki}
\end{figure}

We followed up with continuous improvement analyst Nati in December 2020 as she continued to review and revise documentation practices within Alamo. By then, Alamo's documentation revolved around a format they called \emph{the Wiki} that was conceived as a collaborative document to enable the conduction of projects (see Fig. \ref{fig:wiki}). Each Wiki comprised information, collected in many cases in direct collaboration with the requester, about processes, tools, and best practices concerning each specific project. Nati's main concern back then was reducing the time and effort that were flowing into documentation and, at the same time, finding efficient ways of keeping documentation updated. 

After these interviews with Nati, Pablo, and some other data workers, we got back to Alamo's management with feedback. Among other things, we highlighted the need to reduce the amount of work put on Alamo's team in terms of producing and maintaining documentation and \textcolor{black}{distributing} responsibilities more equitably.  The idea of producing exhaustive metadata for each project was commendable but the price of such an effort was being paid by the data workers.  Moreover, we found that the original idea of the Wikis, namely, that of preserving the knowledge generated in each project and offering more information about the ML supply chain to foster workers' professional growth, had shifted \textcolor{black}{toward} a constant consideration of what clients might want to see reflected in documentation, as Sole, leader of another data labeling team at Alamo, described:

\begin{displayquote}
We need to know what we want to document first.  Who we are documenting for.  That's a challenge at the moment. Because right now I ask myself: who am I doing this for? What for? […] Right now, we're focusing on the clients: what is the client asking for? Because I may want many things, but what is the client interested in? What is the client expecting from us as service providers?
\end{displayquote}

In view of these challenges, we proposed the idea of conducting a series of co-design workshops with Alamo's data workers and managers with the purpose of moving the focus \emph{away from “what clients want” and re-imagining documentation based on the needs of workers.} In June 2021, we started conversations with Alamo.  Negotiation points included participants' compensation and attendance, i.e. how many and which workers would Alamo allow to participate in the workshops. 
We agreed to use the Wiki as the prototype we would discuss and re-imagine at the workshops. After several months of negotiation, the workshops finally took place in October 2021.

\subsubsection{Co-Designing Documentation with Alamo's Workers}
\label{sec:workshops-alamo}
\hspace{1cm}

The workshops with Alamo workers and managers started with a discussion of the Wiki template.  The discussion focused on a recent project from Alamo’s main client, an e-commerce corporation that operates primarily in Latin America. The client instructed Alamo to produce a training dataset with images of false identity cards from three Spanish-speaking Latin American countries. This dataset was later used by the client to train an algorithm to screen IDs. The instructions for workers involved the collection of images of authentic identification cards, their modification to render them invalid, and the labeling of the resulting photos. 

The managers participating in the session mentioned that the main benefit of the Wikis is for data workers to take an interest in existing projects, explore them through the respective Wikis, and use that information for their education and growth. In the words of a project manager:

\begin{displayquote}
We've been working with growth metrics for each person working at Alamo this year. For example, if someone wants to learn programming skills, they can look at the Wiki, see what tools are available, the projects, the languages used, and know what skills to develop.
\end{displayquote}

However, while managers want to encourage workers to have an overview of the undergoing projects at the company, they are also mindful of their clients’ demands in terms of confidentiality and security. For this reason, either much of the information in the documentation lacks details and appears general and superficial, or the access to such information is restrained. Several of the data workers argued that the restrictions and the lack of access to the Wikis as well as difficulties keeping the information updated represented a hurdle for their work:

\begin{displayquote}
One of the limitations is access. That we all know the information is there and can access it. How can we make that happen? Not everyone has access. How can we guarantee that the information is updated? It takes a lot of time to write down the documentation. 
\end{displayquote}

In addition, the data workers asked the managers for trust in accessing complete versions of the documentation, but the managers argued that the information is sensitive and clients often request not to make it available to all. This tension, coupled with the difficulty of keeping documents updated as projects evolve, makes it hard for workers to incorporate feedback in the documentation due to the lack of information. 

Moreover, the workers acknowledged that they were less inclined to use the information on Wikis for reference. For one thing, some workers were not aware that the Wikis were available for them to see other projects and had to be reminded of their presence. For another, the documentation was perceived as “on the machine” and “not close at hand” and, thus, it was not the workers' initial responses to check if relevant information could be found on the Wikis. By the end of day 1, it became evident that there is a discrepancy between how management thinks workers should use the Wikis and the latter’s perceived usefulness vis-à-vis the time and effort required to maintain these documents. 

\begin{figure}[ht]
    \centering
    \includegraphics[scale=0.62]{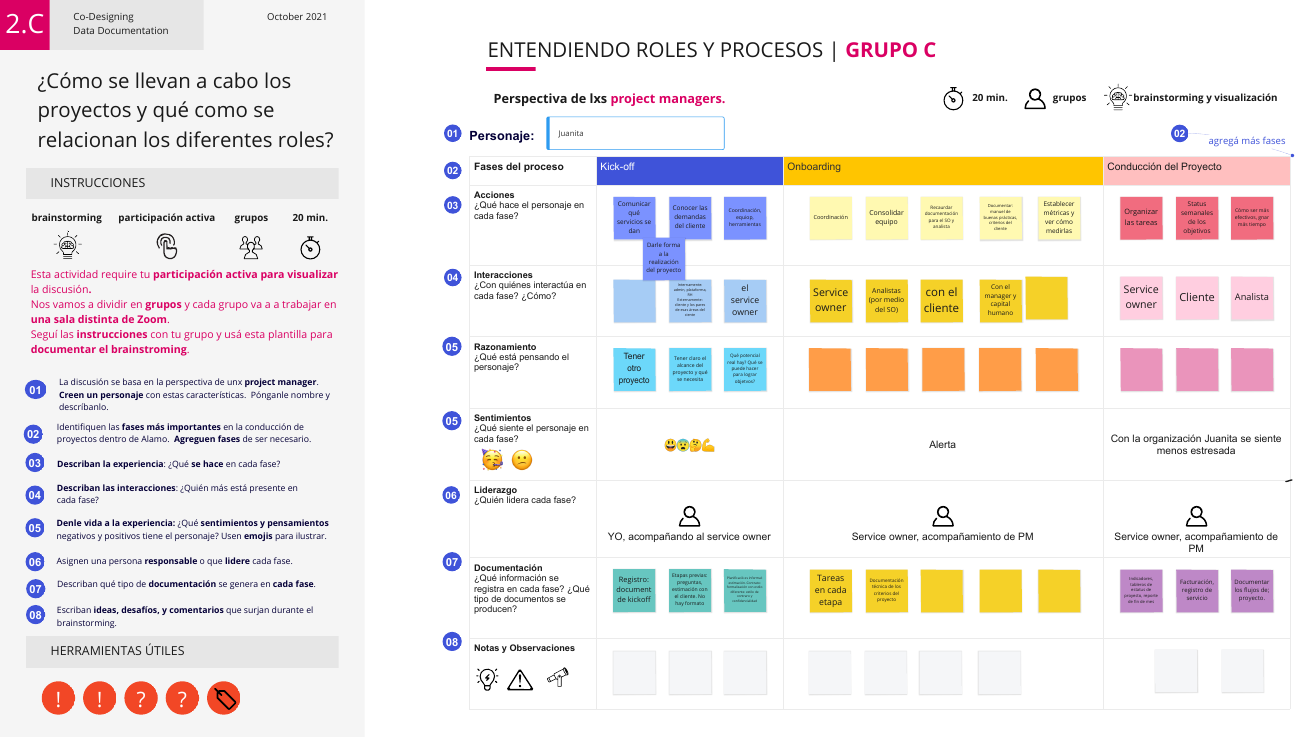}
    \caption{One of the workshop activities aiming to understand roles, workflows, and collaboration within Alamo.}
    \label{fig:userjourney}
\end{figure}

We started Day 2 with a focused discussion of the tensions identified during the previous session: \emph{how to keep the Wikis up to date, how to guarantee access and promote use, how to provide details while keeping confidentiality, and how to integrate documentation practices in existing workflows.}
When discussing how to update the documentation efficiently, one of the data workers pointed out that gathering required information often took weeks due to the organizational structure and dispersed nature of teams.
In this discussion, possible forms of documentation were explored. A former data analyst, now team leader, argued that automating the production of metadata was probably not the best solution because much information exchange and knowledge transfer took place in daily conversations. He argued that \emph{the potential of documentation lies in integrating it into communication processes:} 

\begin{displayquote}
I think that's where the greatest possibility of improving communication lies. To leave at least a record. It does not necessarily have to be so precise to set up a whole process, but a spreadsheet where I document things, an e-mail, a chat. I don't know. There are many ways of communication. I don't know if the best thing is automation, because we are quite beyond automation. Yes, it is a matter of communication.
\end{displayquote}

Finally, the last workshop activity asked workers to imagine possible approaches to the problems discussed earlier in the session. We asked them to think of a documentation process that would guarantee simplicity, accessibility, and integration in workflows. These requirements were derived from the previously conducted interviews.  In their own way, all teams focused on the use of documentation rather than the act of documenting. They all agreed that documentation should be centralized rather than dispersed in different locations and be easy to edit and interpret by any team member. In this sense, three themes were salient. First, the documents should have \emph{better navigation tools for readers.} Some proposed additions include a history of editions (such as in other Wikis like Wikipedia) and a glossary to enable translation between languages and illustrate technology jargon. Second, \emph{documentation should be more legible.} In this sense, participants suggested including visual material, especially in task instructions, such as graphics or video tutorials, and reducing text. Some participants also proposed more interactive documentation. For example, teams can benefit from more precise structures instead of documents without format. Third, participants suggested three ways to \emph{increase participation in the documentation:} making everybody in the organization in charge of documenting, automating the passage between text input and the graphic output, and adding elements of gamification to encourage and measure participation. 

\subsection{Case 2: Action Data}
\label{sec:ActionData}

The Bulgarian company \textit{Action Data} specializes in image data collection, segmentation, and labeling. Its clients are computer vision companies and academic institutions, mostly located in North America and Western Europe. Action Data offers its data workers contractor-based work and the possibility to complete their assignments remotely, with flexible hours. Workers are paid per piece (image/video or annotation), and payment varies according to the project and its difficulty. At the time of our first visit in July 2019, Action Data operated with one manager and two coordinators in salaried positions handling operations and a pool of around 60 freelance data workers. As part of its impact-sourcing mission, Action Data recruited its data workers among refugees from the Middle East who had been granted asylum in Bulgaria. That approach changed in 2021 and now Action Data outsources its projects to partner organizations located in Syria and Iraq. In fact, in November 2021, 90\% of Action Data's projects were conducted by data workers in Syria using a re-intermediation model \cite{Graham2017} in which a third-party organization that we will call \textit{Ranua} recruits the workers while Action Data manages the projects with clients. This model of using a third party to recruit workers has been documented in the crowdsourcing and gig economy sectors, in which some platforms recruit workers to do data work for other platforms \cite{tubaro2020}.

\subsubsection{Investigating Action Data's Documentation Practices}
\label{sec:ADInt}
\hspace{1cm}

In 2019, as we began this investigation, Action Data's projects and workers fluctuated frequently and communication seemed to adapt to the needs of each project. Consequently, Action Data did not have standardized documentation guidelines.
This was critical considering the fact that most of the workers worked remotely after the initial training at the company's office. 

After that first fieldwork period and once we had analyzed the collected data, we provided Action Data with a report summarizing our observations\,---\,most of them related to labor conditions, highlighting their effects on the quality of the datasets produced by the company.  In addition, we provided recommendations regarding the need to educate data workers about the ML pipeline, the urgency of considering labor conditions an essential point in the “ethical AI” discourse at the core of the firm's marketing strategy, and the importance of documenting workflows and practices \cite{kazimzade2020}. The company's management found that our suggestions were useful and actionable. By September 2020 they proceeded to implement some of them, starting with an education program for data workers.  The program included a machine learning course that covered technical and ethical aspects of ML as well as an overview of the ML supply chain.

It was around that time that we invited Action Data to participate in a series of follow-up interviews centered around documentation practices. An important development was the company's implementation of an internal database of all the projects they had conducted since the beginning of their operations. However, some of the information contained in it was considered confidential or sensitive, \textcolor{black}{which} resulted in the documents only being shared with Action Data's management and the respective clients. The data workers did not have access to the project database. We provided feedback on this issue and highlighted the importance of sharing information with the data workers. 

The recently hired operations officer \textcolor{black}{at the time}, Tina, was then put in charge of creating a series of documents targeted \textcolor{black}{at} data workers, that contained guidelines and information about the different projects carried out by Action Data. Apart from our recommendation, two factors contributed to the company's decision to develop these documents. First, Action Data had managed to establish itself in the market of data services for machine learning\,---\,especially data labeling. This means that, by 2020, the company was conducting long-term projects for a more or less steady set of clients, which resulted in the need to keep better records and continually train workers. Second, because of its growth, the company had started outsourcing some requests to partner organizations located in the Middle East. This development revealed the need for clear\,---\,written\,---\,guidelines for a geographically distributed workforce that, in most cases, did not speak English.

\begin{figure}[ht]
    \centering
    \includegraphics[scale=0.40]{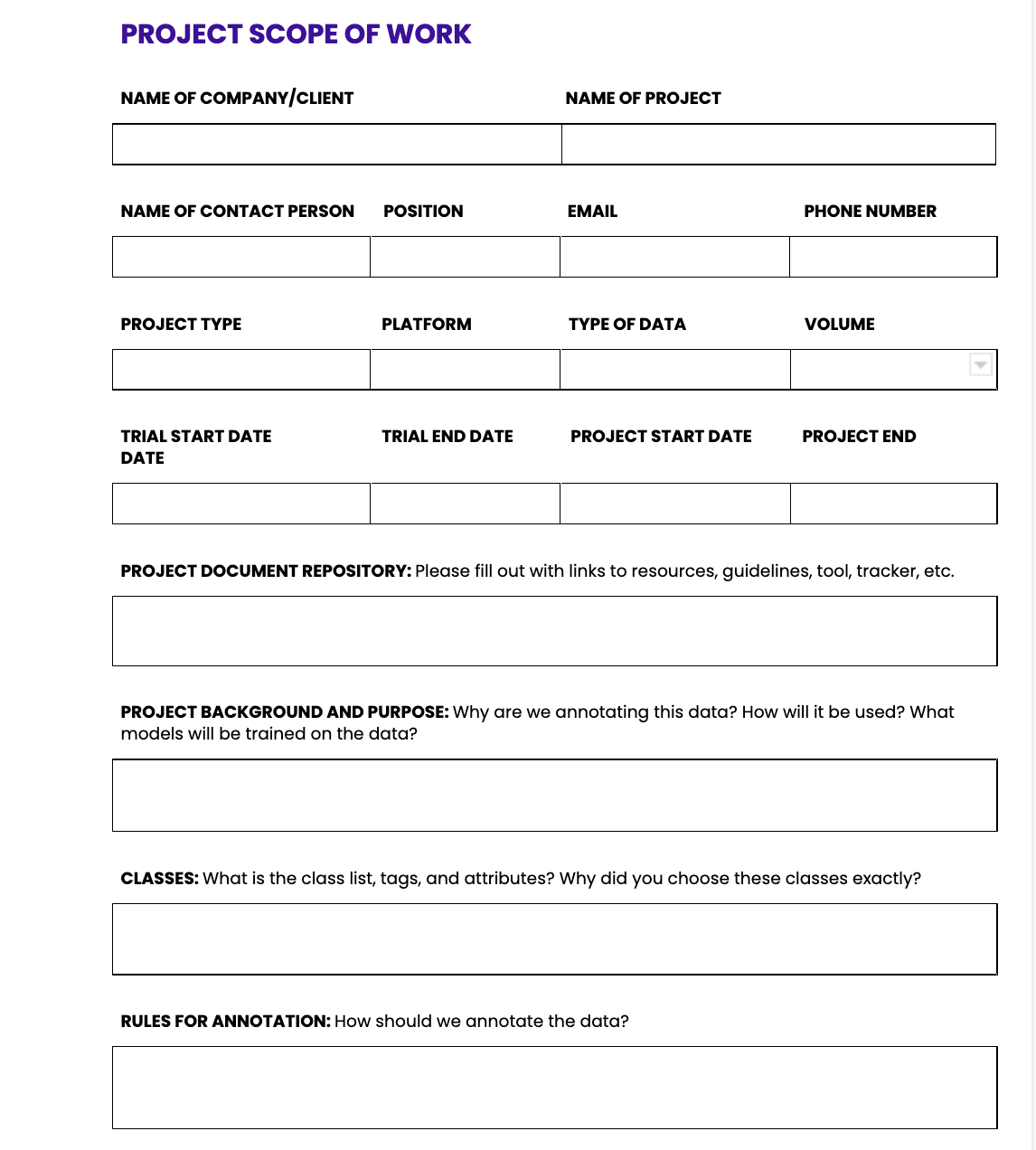}
    \caption{One of the documentation templates prototyped by Action Data: The Scope of Work form (SoW).}
    \label{fig:sow}
\end{figure} 

Since the inclusion of third-party organizations in the conduction of projects and because of the disadvantaged position of service providers vis-à-vis their clients, Action Data's perceived usefulness of documentation started to shift \textcolor{black}{toward} the idea of \emph{inter-organizational accountability} \cite{miceli2020}. The company started to develop a documentation template that was able to reflect its business relationship with other organizations, i.e. clients and partner organizations in the Middle East to which projects were being increasingly outsourced. This form of documentation would make explicit each organization's responsibilities in terms of work, payment, and deadlines and would reflect iterations and updates. Moreover, keeping clear records of the task instructions provided by clients and documenting instructed changes could provide proof that tasks were carried out as instructed, especially in cases where clients are not satisfied with the quality of the service provided or decide to demand more, as Eva explained:

\begin{displayquote}
We also keep the client accountable so that they don't come up with a new requirement or something that we haven't mentioned before. So, documents are also for accountability of us \textcolor{black}{toward} the client as well so that the client can have a document where they can keep track of what the arrangement is and so on beyond our contract.
\end{displayquote}

\textcolor{black}{toward} the beginning of 2021, Action Data began developing three documentation templates. The prototypes were designed as forms \textcolor{black}{where} information would be filled in by Action Data's management. When sending the prototypes to us to request feedback, operations manager Tina explained via email: 

\begin{displayquote}
On the first page is the \emph{SoW} (Scope of Work) that we regularly use in order to scope clients' projects before we start working. On the second one, there is a new idea Eva is piloting for \emph{Dataset Documentation}, so some information we will fill out during and after projects so that the client can use it for transparency purposes if they wish. And then the third page is a \emph{Post Mortem} template that we can use to address issues and challenges after the project is completed.
\end{displayquote}

These three prototypes were examined and interrogated at the co-design workshops conducted with Action Data workers that we describe in the next subsection.

\subsubsection{Co-Designing Documentation with Action Data's Workers}
\label{sec:workshop-AD}
\hspace{1cm}

The co-design workshops took place in three different sessions facilitated over Zoom.

\begin{figure}[ht]
    \centering
    \includegraphics[scale=0.30]{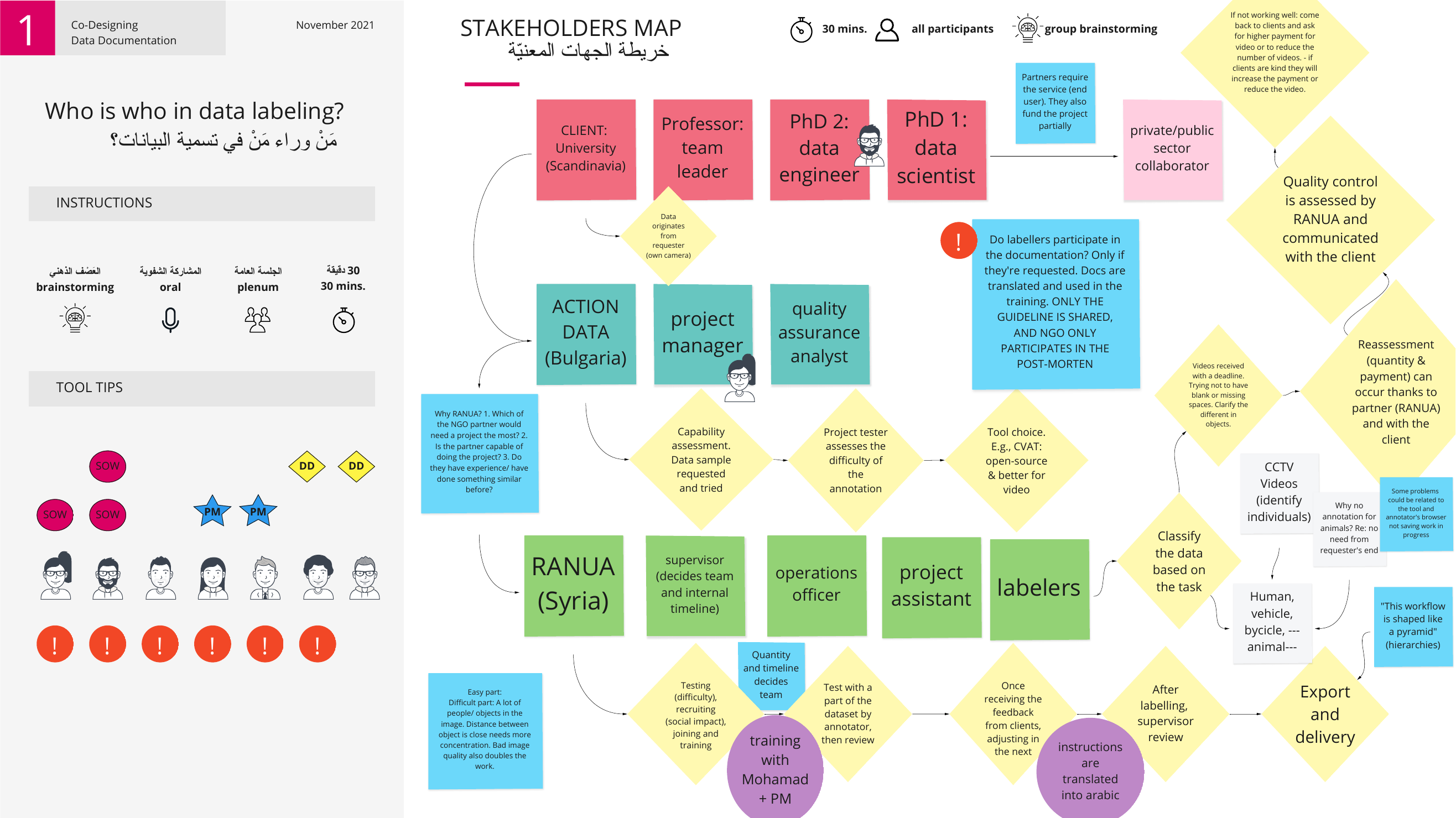}
    \caption{Stakeholder map showing the three groups participating in the second workshop session, the nature of their relationship, and the workflows that characterize it. The map was developed with the workshop participants.  The real names of the organizations and individual actors, originally included in the stakeholder map, have been changed to be included in this paper.}
    \label{fig:stakeholders}
\end{figure}

Day 1 worked as an introduction and focused on getting to know Action Data’s general structure and workflows, and discussing how stakeholders use the three documentation prototypes (“Scope of Work,” “Dataset Documentation,” “Postmortem Report”). To this end, participants mapped the actions and relationships of each of the stakeholders in the data labeling process, from clients to Action Data’s intermediation, to their partner organization Ranua (pseudonym) in Syria, and the Syrian data workers (see Fig. \ref{fig:stakeholders}).  

Day 2 introduced the perspectives of the clients \,---\,three researchers at a Nordic university. They were working on automating a safety system at the harbor front that would release a signal, for instance, if people fall in the water. For this client, the workers in Syria annotated thermal videos from CCTV footage that would serve as training data. The task was coordinated by Action Data and outsourced to Ranua's workers in Syria.

\begin{figure}[ht]
    \centering
    \includegraphics[scale=0.62]{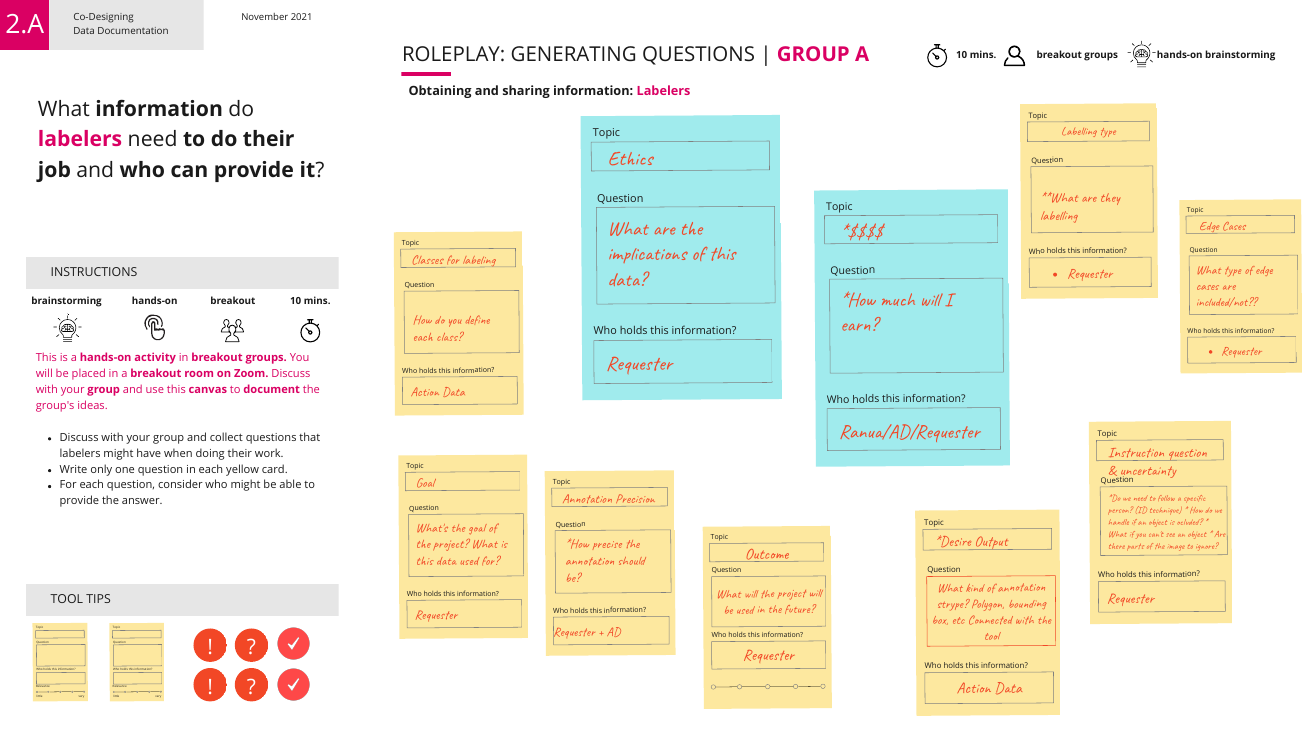}
    \caption{Roleplay activity in breakout groups. Here the requester group had to generate questions (the yellow cards) assuming the role of the data labelers. Afterward, the data labelers provided feedback and added their own questions (in the blue cards.)}
    \label{fig:roleplay}
\end{figure}

The first group activity of the day was a roleplay (see Fig. \ref{fig:roleplay}). We asked each group to reflect on the knowledge required to conduct data-production projects in their assigned roles by filling information cards. Afterward, each team presented \textcolor{black}{their} cards in a discussion round during which each group gave their feedback on how accurately it had been roleplayed by the others. The clients in the role of data workers generated many cards comprising questions about labeling type and classes, edge cases, desired output, and annotation precision (see Fig. \ref{fig:roleplay}). However, the feedback of the actual data workers \textcolor{black}{uncovered} two further issues that the clients generating the questions had not contemplated. First, workers wanted to know how much they would earn. And second, it was important to them to include information about \textcolor{black}{the} ethical implications of the data they had to work with in terms of potential harms to data subjects as well as concerns about the mental health of data workers dealing with violent, offensive, or triggering data. 

In the same discussion, workers considered the extent of annotation precision required an important information type because it defines the difficulty of the task and can have negative effects on the received wages. This requirement is usually negotiated between requesters and managers without the participation of data workers. 
The dispersion and intermediation\,---\,and power\,---\,of actors like Action Data and Ranua also render pricing negotiation inaccessible to workers. These negotiations occur primarily between the client and Action Data with lesser participation of managers from Ranua.

Following these discussions, participants acknowledged that one of the main problems of the documentation prototypes developed by Action Data is that they only concerned managers and clients while ignoring data workers. Moreover, instruction documents for workers are translated into Arabic by Ranua’s management due to workers' limited knowledge of the English language. In this sense, the English-based documentation prototypes proposed by Action Data remain inaccessible to most Syrian workers. 

Finally, the last activity of day 2 prompted participants to imagine new forms of documentation by thinking of its format, structure, information, and the stakeholders involved in its development and access. 
Similar to the Alamo team, Action Data's workers prioritized a view of \emph{documentation as a communication medium.} Mostly, they acknowledged that documentation should allow the co-creation of task instructions with different stakeholders, particularly workers. In the words of one of the Syrian workers:

\begin{displayquote}
Communication and feedback should be the highest priorities. The biggest loser when communication is lacking is the annotators because they spend more effort and time adapting their work each time.
\end{displayquote}

In addition, by increasing communication between different stakeholders, the documentation can potentially \emph{integrate feedback} to make data-production processes clearer by reducing redundant work. Furthermore, it was discussed that a post-mortem report aiming at reflecting on the finished projects allows producing knowledge for future tasks. Such a document could include missing information, met goals, fairness in pricing and wages, the quality of the work environment, difficulties, and learned lessons. 

Finally, on day 3, we hoped to enable a space for data workers to express their opinions and wishes without the presence of managers.
Although all workshop sessions were designed to be about documentation practices, there were some more pressing needs that were the focus of the data workers: to earn a living wage and be able to provide for their families. This way, we opened up the floor to discuss these issues and talk about possible approaches. After the workshops, we helped facilitate the dialogue \cite{bodker_bigissues_2018} between the Syrian data workers and Action Data management.  One of the main issues described by the workers was that \emph{they received the same payment regardless of their effort or time spent on the tasks.} One of the data workers explained the general feeling among her colleagues:

\begin{displayquote}
Some feel that the profit or the earnings they're making after the end of the job are not enough compared to the effort they have put into it. 
\end{displayquote}

Another issue with payment was the lack of room for negotiation by workers about their wages. For example, one practice that workers perceived as unfair was that \emph{management divided the client's payment equally among all team members regardless of individual efforts.} Furthermore, the workers are not consulted in decisions about team composition and cannot influence how many workers will share the payment: 

\begin{displayquote}
When we divide the amount of money that we have received for that certain project, when it's divided by the number of people working on that project, so it turns out to be something completely unfair. 
\end{displayquote}

Tensions were observed during the discussion on the tools used to perform data work. The limitations of tools used by workers require them to share accounts and keep a shared spreadsheet with the number of annotations per task that each worker had done. This makes individual performances visible, accessible, and modifiable to all. One of the workshop participants explained that this was a practical choice due to the diverse working conditions and location of workers and the complexity of supervising and managing their work. He further emphasized mutual trust between the management and workers and among workers. While acknowledging the existence of honest behavior and mutual trust within the team, the other participants did not regard trust as a satisfying justification for the limitations of tools and privacy bridging, and pointed out that the problem was beyond trust issues:

\begin{displayquote}
I get the question of trust, but I believe that some of the issues that have been raised don't really have much to do with trust, but maybe with technical problems or the adequacy of free tools that, you know, make work easier or more difficult. 
\end{displayquote}

For one thing, current tools and arrangements posed difficulties in keeping track of and securing the tasks they individually performed. For another, they did not allow adequate respect for privacy, which could potentially undermine team solidarity, as some participants explained. Therefore, other workers advocated for better tools and arrangements to facilitate work and enhance privacy and security. In this sense, as a potential solution related to documentation efforts, the workers suggested that documentation should reflect how much workers are being paid and the purpose of their work. However, the individual earnings of each worker should remain private.

\subsection{Salient Considerations and Summary of Findings}
\label{sec:FindSum}

Before the workshops, we observed two different orientations in the ways in which the BPOs implemented our feedback to create the documentation prototypes. While Alamo's documentation prototype (\emph{the Wiki}) was oriented \textcolor{black}{toward} the \emph{preservation of knowledge} that could be useful to train (future) workers, Action Data had oriented the design of their documentation templates (especially \emph{the Scope of Work}) \textcolor{black}{toward} keeping \emph{inter-organizational accountability}, i.e., clearly stating clients' expectations and workers’ responsibilities \cite{miceli2021}.

After the workshops, however, we learned that \emph{the needs of data workers at both organizations were not significantly different in terms of what they wanted to see reflected in \textcolor{black}{the} documentation.} The apparent diverging orientations adopted by Action Data and Alamo were based on the corporate needs of each organization because, up until that point, the design process had been piloted by the companies' founders and managers.  
Engaging in hands-on design sessions directly with data workers gave us the possibility to explore their challenges and needs beyond corporate priorities and gave rise to a different set of considerations regarding documentation.


To summarize our findings, we present here five design considerations derived from our long-term engagement at both research sites. Taking steps \textcolor{black}{toward} approaching actionable documentation guidelines, we add concrete questions to illustrate how these considerations could flow into a documentation framework based on data workers' needs.  We present these documentation items as questions because that is how the data workers expressed them at the co-design workshops as we prompted them to re-imagine documentation (see Fig. \ref{fig:co-design}).
For the most part, these observations can be generalized to both BPOs. We added specific notes where differences between the research sites were observed. 

\begin{itemize}
    \item \textbf{The documentation of data production should be collaborative.}  Many actors take part in data production processes and their documentation. The roles can vary according to the context. There are, however, two patterns that can be derived from our findings. First, most of the information about tasks comes from the requesters and is documented by managers. The managers act as gatekeepers, only granting access to information according to the client's confidentiality requests. Second, data workers receive only minimum information that allows them to complete tasks according to \textcolor{black}{the} client's instructions. To break with these patterns, workers must be acknowledged as collaborators and be granted access to information.
    
    $\rightarrow$ \textit{Who is the requester?} The data workers would like to see information about the client, its products, and its relationship history with the data processing company included in the documentation. This information could help workers get an idea of “what to expect” from each project and produce strategies to collaborate more effectively. 
    
    $\rightarrow$ \textit{Who else is working on this project?} Information about team composition is helpful for the coordination and division of labor. Moreover, at Action Data where the project budget is divided among the team members, it is essential for data workers to know how many people will be working on the project because their earnings depend on that.

\hspace{1cm}

    \item \textbf{The documentation of data production should enable communication.} It should transport information that is useful and intelligible to diverse actors and organizations. Language could present a barrier for geographically distributed teams, as reported by Action Data workers. Moreover, communication should not just flow top-down. In this sense, data workers at both sites emphasized the need to enable communication channels for feedback from workers to reach clients.
    
    $\rightarrow$ \textit{What kind of data is this? Where does it come from?} Data workers and managers at both BPOs argued that requesters should communicate key facts about the data, i.e., data provenance, dataset's previous use, and proof of compliance with local regulations, especially in the case of personal data. Moreover, Action Data participants suggested including a warning to inform workers if projects involve violent or offensive material.
    
   $\rightarrow$ \textit{What have we learned from this project?} Participants at both sites argued for the inclusion of a post-mortem report after project completion to reflect upon edge cases, lessons learned, and possible mismatches between time, effort, and payment. Alamo workers considered this to be a useful way to preserve knowledge for their professional growth and for future projects.  Instead, Action Data workers saw such reports as a useful communication channel to provide feedback to requesters, especially on whether pricing and wages were fair vis-a-vis the labor involved in the task.

\hspace{1cm}

    \item \textbf{The documentation of data production should not be seen as a one-time action.} Data production is often messy and iterative. ML datasets evolve and change over time. Therefore, data production requires living, evolving documentation.
    
     $\rightarrow$ \textit{How has this project evolved over time?} Participants at both sites favored the inclusion of an update history log, including the date, time, and name of the person who documented each update. Given the constant negotiation and modification of task instructions and the adaptation of data work projects, a history log could keep important decisions recorded and task instruction and requirements updated. 
     
    $\rightarrow$  \textit{What is the data going to be used for?} Data workers at both sites expressed the need to receive more information about the ML pipeline and dataset maintenance and stewardship. Specifically, the participants wanted documentation to include a description of the ML product that will be trained on the basis of the data produced by them. Regarding the involvement of sensitive or personal data, Action Data workers suggested asking clients for a written commitment that the data will not be used for harmful purposes. In comparison, workers in Alamo pointed to the need \textcolor{black}{for} clear instructions on how to process and store sensitive data to enhance privacy. 
  
  \hspace{1cm}
  
\begin{figure}[ht]
    \centering
    \includegraphics[scale=0.30]{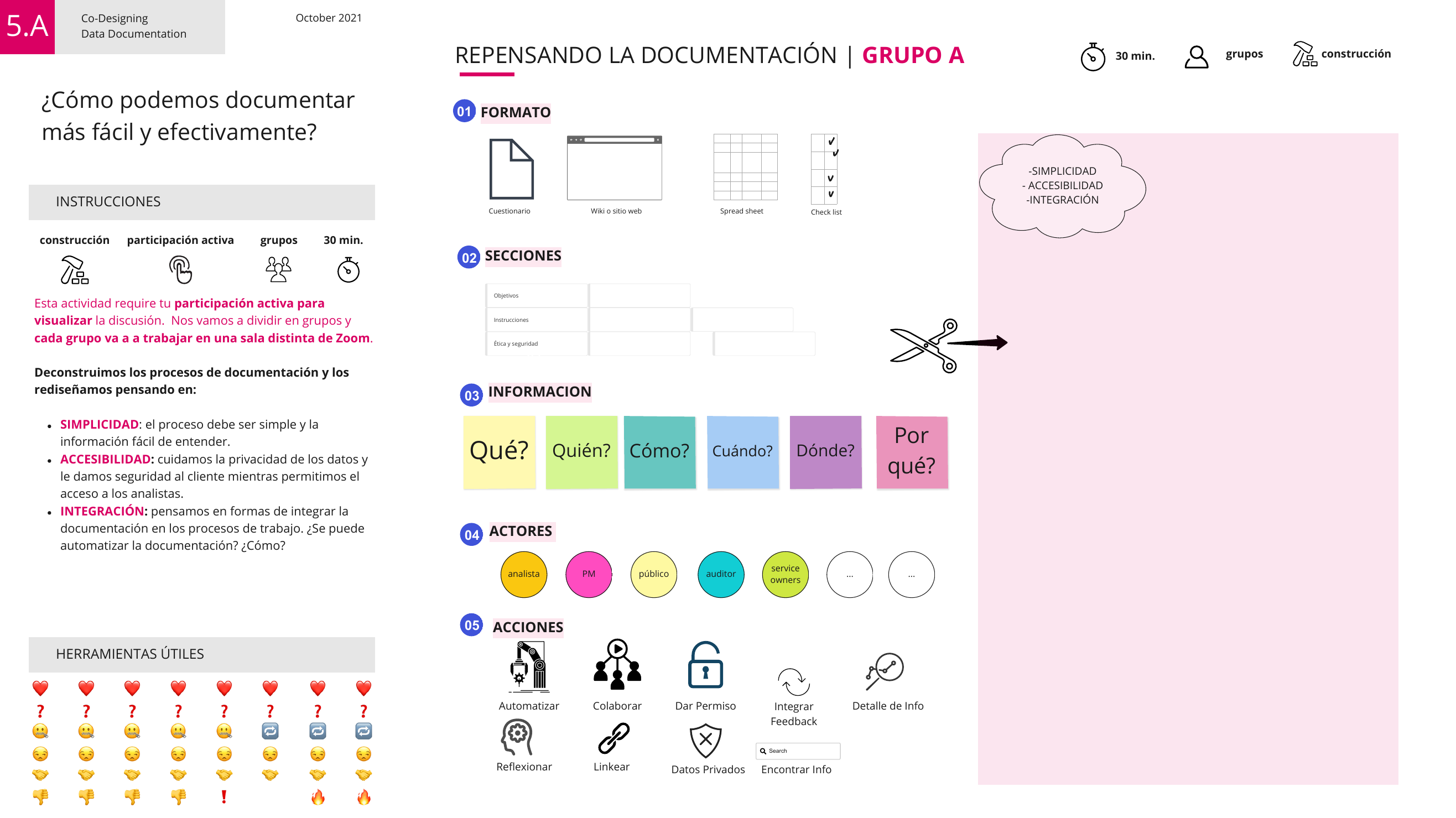}
    \caption{Template for one of the co-design exercises carried out at the workshop series with Action Data and Alamo. The activity prompted data workers to deconstruct and re-imagine documentation practices.}
    \label{fig:co-design}
\end{figure}
  
\hspace{1cm}  
    
    \item \textbf{The documentation of data production should be integrated in existing workflows and routines.} This could help foster a view of documentation as a constitutive part of data production and not as extra work. Action Data workers expressed that documentation could help \textcolor{black}{them} understand projects and their requirements better, which would ease their everyday tasks. Alamo workers mentioned that the preservation of information might come in handy and help them save time in future projects or to train future workers.
    
    $\rightarrow$ \textit{What should be done with the data?} The participants asked to receive a detailed description of tasks, including categories and classes for data annotation and collection, edge cases, and the required quality of annotations. One important point mentioned by Action Data workers was to include information about annotation precision and acceptance and rejection standards which could negatively impact their wages.
   
    $\rightarrow$ \textit{What tools are used?} This includes clear guidelines in terms of communication tools, the software used to work on the data, and \textcolor{black}{the}repository where the data is to be stored. Alamo participants mentioned that the use of certain tools (e.g. Photoshop) has sometimes increased the interest of workers to join certain projects because they wanted to learn that specific software tool.

\hspace{1cm}
    
    \item \textbf{The documentation of data production should be adaptable to stakeholder needs.} We base the design considerations presented in this paper on the needs and wants of data workers and acknowledge that other stakeholders such as requesters, data subjects, or policymakers might have different requirements. Moreover, even if the suggestions and needs of the data workers at both BPOs were similar, some subtle differences could be observed. For instance, the workers at Action Data attached value to the evaluation of project workload and payment details, which would help them assess and decide the extent of commitment to the project, based on the perceived fairness of the wages. Instead, workers in Alamo placed more emphasis on documenting the evolution of projects because the company runs periodical evaluations based on individual and project-based performance metrics.
   
    $\rightarrow$ \textit{How difficult is the task? How \textcolor{black}{labor-intensive} is this data?}  Action Data workers suggested including a “difficulty rating” for workers to quickly estimate the workload and time required for each project. In a similar way, Alamo workers mentioned that an overview of the tools used in each project would help workers understand the difficulty and the knowledge required for specific tasks.
    
   $\rightarrow$ \textit{How much does this project pay each data worker?} Clear information about wages, payment structure, and rules was the most highlighted need at the workshops with Action Data workers because they are paid according to the project and the task. Instead, Alamo workers, who receive a fix\textcolor{black}{ed} salary independently from the project, prioritized information about client expectations and quality standards that would help them complete tasks “according to what each client has in mind,” which could have a positive impact \textcolor{black}{on} workers' performance reviews. 
    
\end{itemize}

\section{Discussion}
\label{sec:dis}


Our findings show that data workers are expected to use the documentation to perform better and learn new skills. However, they receive only the bare minimum information to allow them to complete tasks according to clients’ instructions. Workers’ feedback is rarely considered and only occasionally integrated into task instructions.  In addition, access to documents is not always granted (as reported by Alamo’s workers), and language translations can constitute a barrier (as discussed by Action Data’s workers).
The production of ML datasets involves many iterations and the collaboration of actors working in different organizations that are often geographically distributed.  Moreover, datasets are frequently updated, relabeled, and reused even after being produced.  Therefore, the challenge consists of creating documentation that is able to capture the evolving character of datasets and the intricacies of data work.  To address this challenge, we advocate for a shift of perspective in the development and implementation of data documentation frameworks: \emph{from documenting datasets \textcolor{black}{toward} documenting data production.} 

Data production documentation should travel across multiple stakeholders and facilitate their communication. In this sense, documentation should register iterations in production processes and enable feedback loops. \emph{Documenting data production processes means making collaboration instances and discussions explicit.} By including detailed information about tasks and wages and enabling the feedback of workers, documentation could make precarious labor conditions in outsourced data work explicit and, therefore, contestable.  Previous data documentation initiatives in ML were rooted in the value of transparency and motivated by the need to inform consumers and the general public about dataset characteristics and composition.  Our research shifts the focus away from the notion of transparency and toward promoting reflexivity \cite{dignazio2020, miceli2021, scheuerman2021} in terms of how tasks are laid out and how datasets are produced.  

The acknowledgment that this form of documentation \emph{works as a boundary object} \cite{star1989, bowker1999} can be helpful \textcolor{black}{in approaching} the challenges arising from coordinating and collaborating across organizations \cite{pushkarna2022}. 
In this sense, we recognize two contributions that the study of data documentation as a form of boundary object can bring to the line of research on developing documentation frameworks for datasets.  First, while previous work on data documentation often assumes a single and consistent notion of “dataset producer,” the boundary object notion helps to highlight the multiplicity of involved actors and needs.  This is especially important in data production, as actors that are often globally distributed may have different\,---\,and even divergent\,---\,goals, priorities, and work requirements.  Second, this acknowledgment helps draw more attention to the dynamics between formal standardization and local accommodations, making the local tailoring of documentation practices visible.  As described by \citet{star2010}, we believe that this local adaptation is also a form of invisible work.  Useful as the existing data documentation frameworks are, they remain relatively stable and standardized.  In this regard, making visible how a documentation framework accommodates local specificities and requirements, which can be observed in how it operates as a boundary object, has values for the design of documentation frameworks.


Three aspects identified and explored by \citet{star2010} and \citet{star1989} could be the starting point for valuable considerations regarding the functioning of documentation as a boundary object. The first aspect is \emph{interpretive flexibility}, understood as the adaptability of boundary objects to be interpreted, used, and mobilized differently depending on the communities that use them. 
In our findings, this is manifested in the workers’ view to conceive data documentation primarily as a medium to facilitate the co-creation of task instructions, while managers see documentation as an important site for preserving knowledge and improving performance.  Moreover, companies use documentation to keep each other accountable in the event of discrepancies \cite{miceli2021} and researchers \textcolor{black}{to promote} transparency and reproducibility \cite{gebru2020, geiger2020}.  
The second aspect is the \emph{material and organizational structures} of boundary objects, which arise from the information needs and work requirements of different communities.  Such material and organizational structures could include differential access to information, the ownership of information among workers, and the incorporation of feedback instances. In this sense, the questions included in the previous section reflect the needs and requirements of data workers regarding documentation.
The third aspect is the \emph{tension between the loosely structured common use and more strongly structured local use of boundary objects}.  In view of the ideas expressed by our participants, such tension could be eased through the incorporation of text-based approaches to document processes, visual elements such as video and images to present the documentation, the use of interactive features to retrieve them, and a translation aid to facilitate the process.

While this paper centers on the needs and requirements of data workers, future research could approach the perspectives of other stakeholders, the relationships and communication among them, and how they shape the development of documentation in intra- and inter-organizational contexts through the lens of boundary objects.

\subsection{Design Implications: Documentation \emph{for and with} Data Workers}
\label{sec:DesImp}

Our findings show that designing documentation based on data workers' needs and ideas requires careful consideration of the unique conditions present in machine learning supply chains and in specific data production settings. For instance, the fact that Action Data is in the European Union while Ranua and the data workers are in the Middle East requires higher coordination efforts than data projects at the Argentina-based Alamo.  Similarly, while all workers in Alamo are \textcolor{black}{native} Spanish speakers, in the case of Action Data, English is used as a vehicular second language.  Given that most data workers do not speak English, the language choice renders coordination\,---\,and documentation\,---\,efforts challenging. 
Such variations entail several implications for the design of data production documentation that we list in the following.
 
First, access to information must be guaranteed for documentation to support collaboration.  However, different groups might have diverging needs regarding privacy and security \cite{kroger2021}.  In this sense, trust in collaboration \cite{passi2018a} must be supported by the documentation framework to foster a view of data workers as essential collaborators who need access to information to produce better data \cite{miceli2022}.  Some of the approaches mentioned by our participants included promoting the ownership of information among workers and differentiating personal and group information when restricting access.

Second, requesters could benefit from the early feedback of data workers regarding task design and instructions.  This feedback should also address possible mismatches between payment and effort.  For such an exchange to be possible,  documentation frameworks must enable feedback loops and transparent information about tasks, intended uses of the datasets as well as wages and payment conditions.

Third, to facilitate the integration of documentation in data production workflows, the information input should be simple.  In this sense, the workers favored standardized text-based options such as checklists or multiple-choice questionnaires.
Conversely, visual forms were favored in the documentation presentation.  Some of the participants' ideas included integrating images and short videos to illustrate instruction. 
Moreover, conciseness, good structure, and the integration of a search function and interactive elements were considered crucial to the use of documentation. 

Fourth, the mismatch between a documentation input that should be text-based and a documentation output that should be visually appealing must be considered.  The participants discussed the integration of some form of translation aid, in this case, to turn a concise checklist into a compelling and visual document. 
Translation was also considered in linguistic terms, for instance, in the case of Action Data, where most of the information is documented in one language (English), but the documentation users need it in another (Arabic).  

Fifth, the design of a boundary object such as documentation is not a neutral process \cite{hawkins2017,huvila2011}.  In fact, boundary objects are shaped within power relations that inform how they are understood and enable specific interpretations out of a general interpretative flexibility \cite{trompette2009, hawkins2017}.  The acknowledgment of such power differentials is at the core of our decision to focus on the needs of data workers to re-imagine data documentation and is a critical issue to consider when designing documentation frameworks.

Finally, boundary objects, especially documents, can comprise different, and even divergent, perspectives of involved participants but do not necessarily require reaching a consensus \cite{star1989, star2010}.  Attempts to reach consensus may reinforce hegemonic aspirations and favor the worldviews of certain groups over others \cite{huvila2011, hawkins2017}.  In this regard, any form of documentation based on workers' needs should aim to preserve moments of different and even conflicting viewpoints rather than striving for consensus, as is commonly sought in crowdsourcing tasks.

\subsection{Research Implications: Challenges of Participatory Research}
\label{sec:resimp}

Leaning on the considerations argued by \citet{ledantec_2015}, in what follows, we reflect upon our work's limitations and outline the negotiations and interactions through which we constructed the relationship with our participants. By doing so, we bring to light our position as researchers within the power relations present at the research sites. We believe that the challenges we discuss here are not exclusive to our investigation and can be understood as more general implications for participatory research in CSCW.

First, conducting participatory research with organizations does not mean having full access to observe everything within them. The challenges of negotiating access to corporate sites are not new within CSCW \cite{Wolf_19,passi2020, Wallace_2017, Fiesler_2019}.  In many cases\,---\,including this research\,---\,the participating organizations influence what researchers are allowed to see and to whom they are allowed to talk.  This affects the study sample since, often, individual participants are selected by the organizations' management.  In this sense, our observations were enabled but also constrained by the organizations that hosted us. 

Second, co-designing with partner organizations does not always mean getting to design with the intended \emph{community}.  Often, the gatekeeping tendency of organizations constitutes a barrier for researchers to establish rapport with participants and meaningfully engage in participatory processes with them. 
In our case, even though we had engaged in in-depth interviews with data workers and their views were prioritized in our feedback to the BPOs, the development of the first documentation prototypes was piloted by the companies' founders and managers, and, therefore, based on corporate priorities instead of workers' needs.  After the co-design workshops, we realized that enabling spaces for workers to lead the design process would reveal ideas and requirements that had not surfaced before due to the intermediation of the management.    

Third, including culturally and geographically diverse communities in participatory research and design is as necessary as it is challenging \cite{winschiers2006challenges, mainsah_2014}.  For instance, in our investigation, working across language barriers came with its own set of challenges that, at times, counteracted the dynamic of the workshops. Such cases demand large amounts of researcher adaptability, flexibility, and creativity. 

Finally, being critical while maintaining a relationship with partner organizations requires researchers to balance their commitments \cite{sloane2020}.  We acknowledge that a continuous challenge of our relationship with the BPOs has been finding the right balance between being explicit and critical with our feedback without losing access to the research sites \cite{Wolf_19}.  This also surfaces an issue that many participatory projects face, namely, that of balancing contributions to the research community and to the community that is participating in the design process \cite{ledantec_2015}. Keeping such balance required us to compromise and adapt while remaining reflexive and true to our commitment to the data workers.  


\section{Conclusion}

In the context of industrial machine learning, discretionary decisions around the production of training data remain widely undocumented.  Previous research has addressed this issue by proposing standardized checklists or datasheets to document datasets.  With our work, we seek to expand that field of inquiry \textcolor{black}{toward} a documentation framework that is able to retrieve heterogeneous (and often distributed) contexts of data production. 

For 2.5 years, we conducted a qualitative investigation with two BPO companies, Alamo and Action Data, dedicated to producing ML datasets.  We engaged in a participatory process with data workers at both sites to explore the challenges and possibilities of documentation practices.  The process involved phases of interviewing, data analysis, presentation and discussion of preliminary findings, the collaborative development of prototypes, feedback rounds, and a series of co-design workshops. 

Previous practices at Alamo and Action Data reflect a constant of linear and top-down communication through documentation.  However, as our findings show, designing a documentation framework based on the needs and ideas of data workers involves a different set of considerations.  The data workers prioritized documentation that is able to transport information about tasks, teams, and payment. They also expected documentation to be able to reflect workers' feedback and communicate it back to the requesters.  In this sense, task instructions could be co-created through iterations and communication enabled by documentation.  Our findings include design considerations related to communication patterns and the incorporation of feedback, questions of access and trust, and the differentiation between the creation and use of documentation.

Together with our participants, we have re-imagined data documentation as a process and an artifact that travels among actors and organizations across cultural, social, and professional boundaries and is able to ease the collaboration of geographically distributed stakeholders while preserving the voices of data workers.  Based on this, we have identified and approached several tensions and challenges that emerge from such a form of boundary object.  Breaking with top-down communication and gatekeeping patterns by enabling feedback loops and promoting reflexivity in the documentation process \textcolor{black}{have} been, in this sense, helpful design considerations.



\begin{acks}
Funded by the German Federal Ministry of Education and Research (BMBF) – Nr 16DII113, the International Development Research Centre of Canada, and the Schwartz Reisman Institute for Technology and Society. Our deepest gratitude goes to the workers and organizations that participated in this research. Thanks to Christopher Le Dantec, Marisol Wong-Villacres, and the anonymous reviewers for their comments and suggestions. 
\end{acks}

\bibliographystyle{ACM-Reference-Format}
\bibliography{references}
\end{document}

%% file: participants.tex
\begin{table}[ht]
\caption{\textcolor{black}{Overview of Participants and Research Methods. Some participants were interviewed more than once.}}
\label{tab:participants}
\resizebox{\textwidth}{!}{%
\begin{tabular}{p{0.08\textwidth}p{0.14\textwidth}p{0.27\textwidth}p{0.2\textwidth}p{0.14\textwidth}p{0.17\textwidth}} \hline \hline \small
\textbf{Method} & \multicolumn{1}{c}{\textbf{Organization}} & \multicolumn{1}{c}{\textbf{Description}} & \multicolumn{1}{c}{\textbf{Participants}} & \multicolumn{1}{c}{\textbf{Medium}} & \multicolumn{1}{c}{\textbf{Language}} \\ \hline 
\multirow{7}[23]{*}{{\rotatebox[origin=c]{90}{\small \textbf{SEMI-STRUCTURED INTERVIEWS}}}} & Alamo & BPO in Argentina & 6 data workers & In-Person & Spanish \\
 & Alamo & BPO in Argentina & 2 managers & In-Person & Spanish \\ \cline{2-6}
 & Action Data & BPO in Bulgaria & 9 data workers & In-Person & English with Arabic interpretation \\
 & Action Data & BPO in Bulgaria & 2 managers & In-Person & English \\  \cline{2-6}
 & Alamo's clients & ML Practitioners requesting data work from Alamo & 1 requesters & Virtual & Spanish \\  \cline{2-6}
 & Action Data's clients & ML Practitioners   requesting data work from Action Data & 5 requesters & Virtual & English \\  \cline{2-6}
 & Other Clients & ML Practitioners requesting data   work from other BPOs & 4 requesters & Virtual & English \\ \hline 
\multirow{7}[17]{*}{{\rotatebox[origin=c]{90}{\small \textbf{CO-DESIGN WORKSHOPS}}}} & Action Data & BPO in Bulgaria & 2 managers & \multirow{5}{*}{Virtual} & \multirow{5}{0.17\textwidth}{English with Arabic interpretation} \\
 & Action Data & BPO in Bulgaria & 2 data workers &  &  \\  \cline{2-4}
 & Ranua & Action Data's Syrian partner organization & 1 manager &  &  \\
 & Ranua & Action Data's Syrian partner organization & 13 data workers &  &  \\  \cline{2-4}
 & Action Data's clients & ML Engineers at Scandinavian University & 3 requesters &  &  \\  \cline{2-6}
 & Alamo & BPO in Argentina & 13 data workers & \multirow{2}{*}{Virtual} & \multirow{2}{*}{Spanish} \\
 & Alamo & BPO in Argentina & 4 managers &  & \\ \hline \hline 
\end{tabular}
}
\end{table}

%% file: codebook.tex
\begin{table}[ht]
\caption{Data Analysis: Summary of codes and themes}
\label{tab:codes}
\small
\resizebox{\textwidth}{!}{%
\begin{tabular}{p{0.25\textwidth}p{0.3\textwidth}p{0.45\textwidth}}
\hline \hline
\multicolumn{1}{c}{\textbf{Domain Summaries}}          & \multicolumn{1}{c}{\textbf{Codes}} & \textbf{Themes}                                                                                                                                    \\ \hline
\multirow{7}{0.25\textwidth}{FUNCTION OF DOCUMENTATION}             & Standardization                    & \multirow{13}{0.45\textwidth}{\textbf{DOCUMENTATION AS COMMUNICATION ENABLER ACROSS BOUNDARIES}}                                                                \\ \cline{2-2}
                                                       & Foster Workers Growth              &                                                                                                                                                    \\ \cline{2-2}
                                                       & Quality Control \& Metrics         &                                                                                                                                                    \\ \cline{2-2}
                                                       & Work Enabler                       &                                                                                                                                                    \\ \cline{2-2}
                                                       & Inter-org. Accountability &                                                                                                                                                    \\ \cline{2-2}
                                                       & Knowledge Preservation              &                                                                                                                                                    \\ \cline{2-2}
                                                       & Communication Medium               &                                                                                                                                                    \\ \cline{1-2}
\multirow{6}{0.25\textwidth}{ROLES AND COLLABORATION}               & Auditor                            &                                                                                                                                                    \\ \cline{2-2}
                                                       & Managers                           &                                                                                                                                                    \\ \cline{2-2}
                                                       & Service Owners                     &                                                                                                                                                    \\ \cline{2-2}
                                                       & Labelers                           &                                                                                                                                                    \\ \cline{2-2}
                                                       & BPO's Teams                        &                                                                                                                                                    \\ \cline{2-2}
                                                       & Clients                            &                                                                                                                                                    \\ \hline
\multirow{5}{0.25\textwidth}{DOCUMENTATION FORMATS}                 & Wiki                               & \multirow{10}{0.45\textwidth}{\textbf{HOW DOCUMENTATION CAN SUPPORT WORKERS’ AGENCY}}                                                                            \\ \cline{2-2}
                                                       & Kick-off Document                  &                                                                                                                                                    \\ \cline{2-2}
                                                       & Scope of Work (SoW)                &                                                                                                                                                    \\ \cline{2-2}
                                                       & Dataset Documentation              &                                                                                                                                                    \\ \cline{2-2}
                                                       & Postmortem Report                  &                                                                                                                                                    \\ \cline{1-2}
\multirow{5}{0.25\textwidth}{WORK ETHIC AND LABOR CONDITIONS}      & Impact and Responsibility          &                                                                                                                                                    \\ \cline{2-2}
                                                       & Trust                              &                                                                                                                                                    \\ \cline{2-2}
                                                       & Security                           &                                                                                                                                                    \\ \cline{2-2}
                                                       & Labor Conditions                   &                                                                                                                                                    \\ \cline{2-2}
                                                       & Confidentiality                    &                                                                                                                                                    \\ \hline
\multirow{6}{0.25\textwidth}{LIMITATIONS OF CURRENT DOC. PRACTICES} & Integration in Workflows           & \multirow{6}{0.45\textwidth}{\textbf{CHALLENGES DESIGNING FOR WORKERS IN CORPORATE ENVIRONMENTS}}                                                               \\ \cline{2-2}
                                                       & Use of Documentation               &                                                                                                                                                    \\ \cline{2-2}
                                                       & Lack of Information                &                                                                                                                                                    \\ \cline{2-2}
                                                       & Access                             &                                                                                                                                                    \\ \cline{2-2}
                                                       & Feedback                           &                                                                                                                                                    \\ \cline{2-2}
                                                       & Updates                            &                                                                                                                                                    \\ \hline
\multirow{8}{0.25\textwidth}{POSSIBLE DESIGN SOLUTIONS}             & Ownership                          & \multirow{8}{0.45\textwidth}{\textbf{DESIGN IDEAS EMERGED FROM THE WORKSHOPS}}                                                                                  \\ \cline{2-2}
                                                       & Synthesis /Simplicity              &                                                                                                                                                    \\ \cline{2-2}
                                                       & Interactivity                      &                                                                                                                                                    \\ \cline{2-2}
                                                       & Information Type                   &                                                                                                                                                    \\ \cline{2-2}
                                                       & Integration                        &                                                                                                                                                    \\ \cline{2-2}
                                                       & Centralizing                       &                                                                                                                                                    \\ \cline{2-2}
                                                       & Automation                         &                                                                                                                                                    \\ \cline{2-2}
                                                       & Visualization                      &                                                                                                                                                    \\ \hline
\multirow{4}{0.25\textwidth}{LESSONS}                             & Lessons from Fieldwork           & \multirow{4}{0.45\textwidth}{\textbf{WHAT PARTICIPANTS \& ORGANIZERS LEARNED THROUGH THE CO-DESIGN PROCESS}} \\ \cline{2-2}
                                                       & Lessons from our Feedback        &                                                                                                                                                    \\ \cline{2-2}
                                                       & Lessons from the Interviews      &                                                                                                                                                    \\ \cline{2-2}
                                                       & Lessons from the Workshops       &                                                                                                                                                    \\ \hline \hline
\end{tabular}
}
\end{table}